\documentclass[usegraphicx,usedcolumn,usenatbib]{mn2e}
\usepackage{astbibref}
\usepackage{color,epsfig,amssymb,longtable}
\usepackage{array,colortbl}
\def\HII{H{\sc ii}}

\begin{document}
\title[Accuracy of abundance determinations of \HII\ galaxies]{The temperature and
  ionization structure of the emitting gas in \HII\ galaxies: Implications for
  the accuracy of abundance determinations.}
\author[G. H\"agele et al.]
{Guillermo~F.~H\"agele$^{1}$\thanks{PhD fellow of Ministerio de Educaci\'on y
    Ciencia, Spain; guille.hagele@uam.es},
Enrique~P{\'e}rez-Montero$^{1}$, \'Angeles~I.~D\'{\i}az$^{1}$,  
\newauthor Elena Terlevich$^{2}$ and Roberto Terlevich$^{2}$\thanks{Research
  Affiliate at IoA}
\\
$^{1}$ Departamento de F\'{\i}sica Te\'orica, C-XI, Universidad Aut\'onoma de
Madrid, 28049 Madrid, Spain\\ 
$^{2}$ INAOE, Tonantzintla, Apdo. Postal 51, 72000 Puebla, M\'exico\\ }
\date{Accepted 
      Received ;
      in original form }
\pagerange{\pageref{firstpage}--\pageref{lastpage}}
\pubyear{2005}
\maketitle
\begin{abstract}

We propose a methodology to perform a self-consistent analysis of the
physical properties of the emitting gas of HII galaxies adequate to the data
that can be obtained with the XXI century technology. This methodology
requires the production and calibration of empirical relations between the
different line temperatures that should superseed currently used ones based on
very simple, and poorly tested,  photo-ionization model sequences. 

As a first step to reach these goals we have obtained simultaneous blue to far
red longslit spectra with the William Herschel Telescope (WHT) of three compact
\HII\ galaxies selected from the  Sloan Digital Sky Survey (SDSS) Data Release 2
(DR2) spectral catalog using the INAOE Virtual Observatory superserver. Our
spectra cover the range from 3200 to 10500\,\AA, including the Balmer jump, the
[O{\sc ii}]\,$\lambda\lambda$\,3727,29\,\AA\ lines, the [S{\sc
    iii}]\,$\lambda\lambda$\,9069,9532\,\AA\ doublet as well as various weak
auroral lines such as [O{\sc iii}]\,$\lambda$\,4363\,\AA\ and [S{\sc
    iii}]\,$\lambda$\,6312\,\AA.

For the three objects we have measured at least four line temperatures: T([O{\sc
    iii}]), T([S{\sc iii}]), T([O{\sc ii}]) and T([S{\sc ii}]) and the Balmer
continuum temperature T(Bac). These measurements and a careful and realistic 
treatment of the observational errors yield total oxygen abundances with
accuracies between 5 and 9\%. These accuracies are expected to 
improve as better calibrations based on more precise measurements, both on 
electron temperatures and  densities, are produced. 

We have compared our obtained spectra with those downloaded from the SDSS DR3
finding a satisfactory agreement. The analysis of these spectra yields values of
line temperatures and elemental ionic and total abundances which are in general
agreement with those derived from the WHT spectra, although for most quantities
they can only be taken as estimates since, due to the lack of direct
measurements of the required lines, theoretical models had to be used whose
uncertainties are impossible to quantify.

The ionization structure found for the observed objects from the
O$^{+}$/O$^{2+}$ and S$^{+}$/S$^{2+}$ ratios points to high values of the
ionizing radiation, as traced by the values of the ``softness parameter" 
$\eta$
which is less than one for the three objects. The use of line temperatures
derived from T([O{\sc iii}]) based on current photoionization models
yield for the two highest excitation objects, much higher values of
$\eta$ which would imply lower ionizing temperatures. This is however
inconsistent with the ionization structure as probed by the measured emission
line intensities.  

Finally, we have measured the T(Bac) for the three
observed objects and derived temperature fluctuations.
Only for one of the objects, the temperature
fluctuation is significant and could lead to higher oxygen abundances by about
0.20\,dex.

\end{abstract}


\section{Introduction}

\HII\ galaxies are low mass irregular galaxies with, at least, a recent episode
of  violent star formation \citep*{1985MNRAS.216..255M,1985RMxAA..11...91M}
concentrated in a few parsecs close to their cores. The ionizing fluxes
originated by these young massive stars dominate the light of this subclass of
Blue Compact Dwarf galaxies (BCDs) which show  emission line spectra very
similar to those of giant extragalactic  \HII\ regions \citep[GEHRs;
][]{1970ApJ...162L.155S,1980ApJ...240...41F}. Therefore, by applying the same
measurement techniques as for \HII\ regions, we can derive the temperatures,
densities and chemical composition of  the interstellar gas in this type of
generally  metal-deficient galaxies. In some cases, it is possible to detect in
these objects, intermediate-to-old stellar populations which have a more uniform
spatial distribution than the bright and young stellar  populations associated
with the ionizing clusters \citep{1998ApJ...493L..23S}. This older population
produces a characteristic spectrum with absorption features which mainly affect
the hydrogen recombination emission lines \citep{1988MNRAS.231...57D}, that is
the Balmer and Paschen series in the spectral range of interest. In some cases,
the underlying stellar absorptions can severely affect the ratios of H{\sc i}
line pairs and hence the determination of the reddening constant
(C(H$\beta$)). They must therefore be measured with special care (see discussion
in \S \ref{measure}).

A considerable number of the {blue} objects observed at intermediate
redshifts seem to have properties {(mass, R$_e$, velocity width of  the emission
lines) similar to local \HII\ galaxies}  
\citep{1994ApJ...427L...9K,1995ApJ...440L..49K,1996ApJ...460L...5G,1998ApJ...495L..13G}. 
{In particular, those with $\sigma$\,$<$\,65\,km\,s$^{-1}$ follow the same
$\sigma - L_B\ {\rm and}\ L_{H\beta}$ relation as seen in \HII\ galaxies
\citep{2000MNRAS.311..629M,2003ApJ...598..858M,2003RMxAC..16..213T,2005MNRAS.356.1117S}.}
{Similar conclusion is drawn from recent studies on Lyman Break galaxies that
also suggest that strong narrow emission line galaxies  might have been 
very common in the past
\citep[e.g.][]{2000ApJ...528...96P,2001ApJ...554..981P,2001A&A...379..393E}. 
To detect possible evolutionary effects like systematic 
differences in their chemical composition, accurate and reliable methods
for abundance determination are needed.}

This is usually done by combining photoionization model results and observed
emission line intensity ratios.  {There are  several major problems with this
approach that limit the confidence of  present results.}
Among them: the effect of temperature structure in multiple-zone models
\citep{2003MNRAS.346..105P}; the presence of temperature fluctuations across the
nebula \citep{2003ApJ...584..735P};  collisional and density effects on ion
temperatures (Luridiana et al.\ 1999, P\'erez-Montero \& D\'{\i}az 2003); the
presence of neutral zones affecting the calculation of ionization correction
factors (ICFs; \nocite{2002ApJ...565..668P} Peimbert et al.\ 2002); the
ionization structure not adequately reproduced by current models
\citep{2003MNRAS.346..105P}; ionization vs.~matter bounded zones, affecting the
low ionization lines formed in the outer parts of the ionized regions
\citep*{2002MNRAS.329..315C}. 
\nocite{1999ApJ...527..110L}
\nocite{2005MNRAS.361.1063P}
On the other hand, the understanding of the age and evolutionary state of \HII\
galaxies require the use of self-consistent models for the ionizing stars and
the ionized gas. However, model computed evolutionary sequences show important
differences with observations \citep{2003A&A...397...71S}, including: (a) He{\sc
  ii} is too strong in a substantial number of objects as compared to model
predictions; (b) [O{\sc iii}]/H$\beta$ vs.\ [O{\sc ii}]/H$\beta$ and [O{\sc
    iii}]/H$\beta$  vs.~[O{\sc i}]/H$\beta$ are not  well reproduced by
evolutionary model sequences in the sense that predicted collisionally excited
lines are too weak compared to observations; (c) there is a large spread in the
[N{\sc ii}]/[O{\sc ii}] values (more than an order of magnitude) for galaxies
with the same value of ([O{\sc ii}]+[O{\sc iii}])/H$\beta$ in the metallicity
range from 8 to 8.4 (see e.g. P\'erez-Montero \& D\'iaz 2005). 
\nocite{2002AJ....123..485S}

\begin{table*}
\centering
\caption[]{Journal of observations.  Right ascension, declination and
  redshift were obtained from the SDSS.}
\label{jour}
\begin{tabular} {l c r r c c}
\hline
 \multicolumn{1}{c}{Object  ID}   &  spSpec SDSS   & \multicolumn{1}{c}{RA} & \multicolumn{1}{c}{Dec} &  z  &    Exposure (s) \\
\hline
SDSS J002101.03+005248.1 & spSpec-51900-0390-445  &  5.254297 & 0.880038 &  0.098 & 1 $\times$ 1200 + 2 $\times$ 2400\\
SDSS J003218.60+150014.2 & spSpec-51817-0418-302  &  8.077479 & 15.003949 &  0.018 & 1 $\times$ 1200 + 2 $\times$ 2400\\
SDSS J162410.11-002202.5  & spSpec-52000-0364-187  &  246.042122 & -0.367378 &  0.031 & 1 $\times$ 1200 + 3 $\times$ 1800\\
\hline
\end{tabular}
\end{table*}

\begin{table*}
\centering
\caption[]{SDSS photometric magnitudes obtained using the DR4 explore tools.}
\label{obj}
\begin{tabular} {l c r r c c}
\hline
 \multicolumn{1}{c}{Object  ID}   & \multicolumn{1}{c}{u} & \multicolumn{1}{c}{g} & \multicolumn{1}{c}{r} & \multicolumn{1}{c}{i}   &    \multicolumn{1}{c}{z} \\
\hline
SDSS J002101.03+005248.1 & 17.56 & 17.35 & 17.51 & 16.98 & 17.45  \\
SDSS J003218.60+150014.2 & 17.04 & 16.49 & 16.53 & 16.74 & 16.65  \\
SDSS J162410.11-002202.5  & 17.07 & 16.46 & 16.91 & 16.80 & 16.74  \\
\hline
\end{tabular}
\end{table*}

Substantial progress toward solving the problems listed above has to come from
the accurate measurement of weak emission lines  which will allow to derive 
[O{\sc ii}], [S{\sc ii}] and [S{\sc iii}] temperatures and densities allowing to
constrain the ionization structure as well as Balmer and Paschen discontinuities
which will provide crucial information about the actual values of temperature
fluctuations. It is possible that these fluctuations produce the observed
differences between the abundances relative to hydrogen derived from
recombination lines (RLs) and collisionally excited lines (CELs) when a constant
electron temperature is assumed
\citep{1967ApJ...150..825P,1969BOTT....5....3P,1971BOTT....6...29P}. { These
discrepancies have been observed in a good sample of objects, such as galactic
\HII\ regions \citep[e.g.][and references
therein]{2004MNRAS.355..229E,2005MNRAS.362..301G,2006MNRAS.368..253G}, \HII\
regions in the Magellanic Clouds \citep[e.g.][and references
therein]{2000ApJ...541..688P,2003ApJ...584..735P,2003MNRAS.338..687T},
extragalactic \HII\ regions \citep[e.g.][and references
therein]{2003RMxAC..16..113P,2005ApJ...634.1056P,2006astro.ph..3134G} and
planetary nebulae \citep[e.g.][and references
therein]{2002MNRAS.334..777R,2005MNRAS.362..424W,2006MNRAS.tmp..404L,2006astro.ph..5082L,2006astro.ph..5595P}.
Likewise, there are relatively recent theoretical works that study the possible
causes of these discrepancies in abundance determinations using photoionization
models of different complexity \citep[e.g.][and references
therein]{2005A&A...434..507S,2005A&A...444..723J,2005MNRAS.364..687T}.}

Unfortunately most of the available starburst and \HII\ 
galaxy spectra have only a restricted wavelength  range (usually from about 3600
to 7000 \AA) and do not have the adequate S/N to accurately measure the
intensities of the weak diagnostic emission lines. Even the Sloan Digital Sky
Survey (SDSS) spectra (Stoughton et al.\ 2002) do not cover simultaneously the
[O{\sc ii}]\,$\lambda\lambda$\,3727,29 and the [S{\sc iii}]\,$\lambda$\,9069\,\AA\
lines.
We have therefore undertaken a project with the aim of obtaining a database of
top quality line ratios { for a sample that includes the best objects for the
task. The data is collected using exclusively two arm  spectrographs in order to
guarantee both high quality spectrophotometry in the whole spectral range from
3500 to 10500\,\AA\ and good spectral and spatial resolution.} In this way we
will be in a  position to vastly improve constraints on the photoionization
models including the  mapping of the ionization structure and the measurement of
temperature fluctuations about which very little is known.

In this first work we present observations of three \HII\ galaxies selected from
the SDSS. Details regarding the selection of the objects as well as the
observations are given in Section 2. Section 3 presents the results including
line measuring techniques. The methodology for the derivation of gaseous
physical conditions and elemental abundances is presented in Sections 4 and 5
respectively. The discussion of our results, including a detailed comparison
with the SDSS data is presented in Section 6. Finally, Section 7 summarizes the
main conclusions of our work.

\section{Observations and data reduction}
\label{Obs}

\nocite{2006ApJS..162...38A}
\nocite{2004AJ....128..502A}
\nocite{2005AJ....129.1755A}
 
\subsection{Object selection}

SDSS constitutes a great base for statistical studies of the properties of galaxies. At
this moment, the Fourth Data Release\footnote{http://www.sdss.org/dr4/} (DR4),
the last one up to now, contains five band photometric data for  about
18$\times$10$^7$  objects and more than 6.7$\times$10$^5$ spectra of galaxies,
quasars and stars (Adelman-McCarthy et al.\ 2006). The
spectroscopic data have a resolution (R) of 1800-2100 covering a spectral range from
3800 to 9200\,\AA, with a single 3\,arcsec diameter aperture. The SDSS data were
reduced and flux-calibrated using automatic pipelines. However, when we selected our
objects on July 2004, the DR2\footnote{http://www.sdss.org/dr2/} was just
available. Both data releases, DR2 and DR4, contain the same type of objects
observed in the 
same five photometric bands and using the same spectroscopic configuration. 
DR2 has photometric data for over 88 million unique objects and about
3.7$\times$10$^5$ spectra (Abazajian et al.\ 2004). All the objects belonging
to one data release are also included in the next ones, but some objects could have been
re-calibrated or re-observed somehow. 
Using the implementation of the SDSS database in the INAOE Virtual Observatory
superserver\footnote{http://ov.inaoep.mx/}, we have selected from the whole SDSS
DR2 { the brightest nearby narrow emission line galaxies with very strong
lines. Our selection parameters were: \par
Equivalent width of H$\alpha\,>$\,50\,\AA,\par
1.2\,$<$\,$\sigma$(H$\alpha$)\,$<$\,7\,\AA, \par
redshift , {\em z} $<$ 0.2 and \par
H$\alpha$ flux, F(H$\alpha$)\,$>$\,4\,$\times$\,10$^{-14}$\,erg\,cm$^{-2}$\,s$^{-1}$\,\AA$^{-1}$.

This preliminary list was then processed using BPT
\nocite{1981PASP...93....5B}(Baldwin, Phillips \& Terlevich 1981) diagnostic
diagrams to remove AGN-like objects. 

The final list consists of about 200 bonafide bright \HII\ galaxies. They show
spectral properties indicating a wide range of gaseous abundances and ages of
the underlying stellar populations \citep{jesustesis}. 

From this list, the final set was selected by further restricting the sample 
to the largest H$\alpha$ flux and highest S/N objects.

From the 10 brightest of the final set we selected three HII galaxies to be
observed in our one single night observing run. This} final selection was 
made based on the relative positions of the sources in the sky allowing to
optimize the observing time. The journal of observations is given in table
\ref{jour} and the photometric characteristics of the objects are listed in
table \ref{obj}. Subsequently we have used the explore
tool\footnote{http://cas.sdss.org/astro/en/tools/explore/} implemented in the
DR3\footnote{http://www.sdss.org/dr3/} (Abazajian et al.\ 2005), which was
available at the time of analysis, to extract again 
the three object SDSS spectra  for comparison purposes.


\begin{table}
\centering
\caption[]{WHT instrumental configuration}
\label{config}
\begin{tabular} {l c c c c}
\hline
 & Spectral range  &       Disp.             & FWHM & Spatial res.         \\
 &      (\AA)             & (\AA\,px$^{-1}$) & (\AA)    & (\arcsec\,px$^{-1}$) \\
\hline
blue & 3200-5700   &       0.86              &  2.5      &   0.2                \\
red  & 5500-10550  &       1.64             &  4.8      &   0.2                \\
\hline
\end{tabular}
\end{table}

\subsection{Observational details}

The blue and red spectra were obtained simultaneously using the ISIS double beam
spectrograph mounted on the 4.2m William Herschel Telescope (WHT) of the Isaac
Newton Group (ING) at the Roque de los Muchachos Observatory, on the Spanish
island of La Palma. They were acquired on July the 18th 2004 { during one
single night observing run and under photometric conditions.} EEV12 and Marconi2
detectors were attached to the blue and red arms of the spectrograph,
respectively. The R300B grating was used in the blue covering the wavelength
range 3200-5700\,\AA\ (centered at $\lambda_c$\,=\,4450\,\AA), giving a spectral
dispersion of 0.86\,\AA\,pixel$^{-1}$. On the red arm, the R158R grating was
mounted  providing a spectral range from 5500 to 10550\,\AA\
($\lambda_c$\,=\,8025\,\AA) and a spectral dispersion of
1.64\,\AA\,pixel$^{-1}$. In order to reduce the readout noise of our images we
have taken the observations with the `SLOW' CCD speed. The pixel size for this
set-up configuration is 0.2 arcsec for both spectral ranges.  The slit width
was $\sim$0.5 arcsec, which, combined with the spectral dispersions, yielded
spectral resolutions of about 2.5 and 4.8\,\AA\ FWHM in the blue and red arms 
respectively. All observations were made at paralactic angle to avoid effects of
differential refraction in the UV. The instrumental configuration, summarized in
table \ref{config} was planned in order to cover the whole spectrum from 3200 to
10550\,\AA\  providing at the same time a moderate spectral resolution.   This
guarantees the simultaneous measurement  of the Balmer discontinuity and the
nebular lines of [O{\sc ii}]\,$\lambda\lambda$\,3727,29 and [S{\sc
    iii}]\,$\lambda\lambda$\,9069,9532\,\AA\  at both ends of the spectrum, in
the very same region of the galaxy. A good signal-to-noise ratio was also
required to allow the detection and measurement of weak lines such as  [O{\sc
    iii}]\,$\lambda$\,4363, [S{\sc ii}]\,$\lambda\lambda$\,4068, 6717 and 6731,
and [S{\sc iii}]\,$\lambda$\,6312. 

Several bias and sky flat field frames were taken at the beginning and at the end
of each night in both arms. In addition, two lamp flat fields and one calibration
lamp exposure were performed at each telescope position. The calibration lamp
used was CuNe+CuAr. The images were processed and analyzed with
IRAF\footnote{IRAF: the Image Reduction and Analysis Facility is distributed by
  the National Optical Astronomy Observatories, which is operated by the
  Association of Universities for Research in Astronomy, In. (AURA) under
  cooperative agreement with the National Science Foundation (NSF).} routines in 
the usual manner. The procedure includes
the removal of cosmic rays, bias substraction, division by a normalized flat
field and wavelength calibration. Typical wavelength fits were performed using 30-35
lines in the blue and 20-25 lines in the red and polynomials of second to
third order. These fits have been done at 117 different locations along the slit
in each arm (beam size of 10 pixels) obtaining rms residuals between
$\sim$0.1 and $\sim$0.2\,pix.  

\begin{figure*}
\includegraphics[width=.31\textwidth,height=.63\textwidth,angle=270]{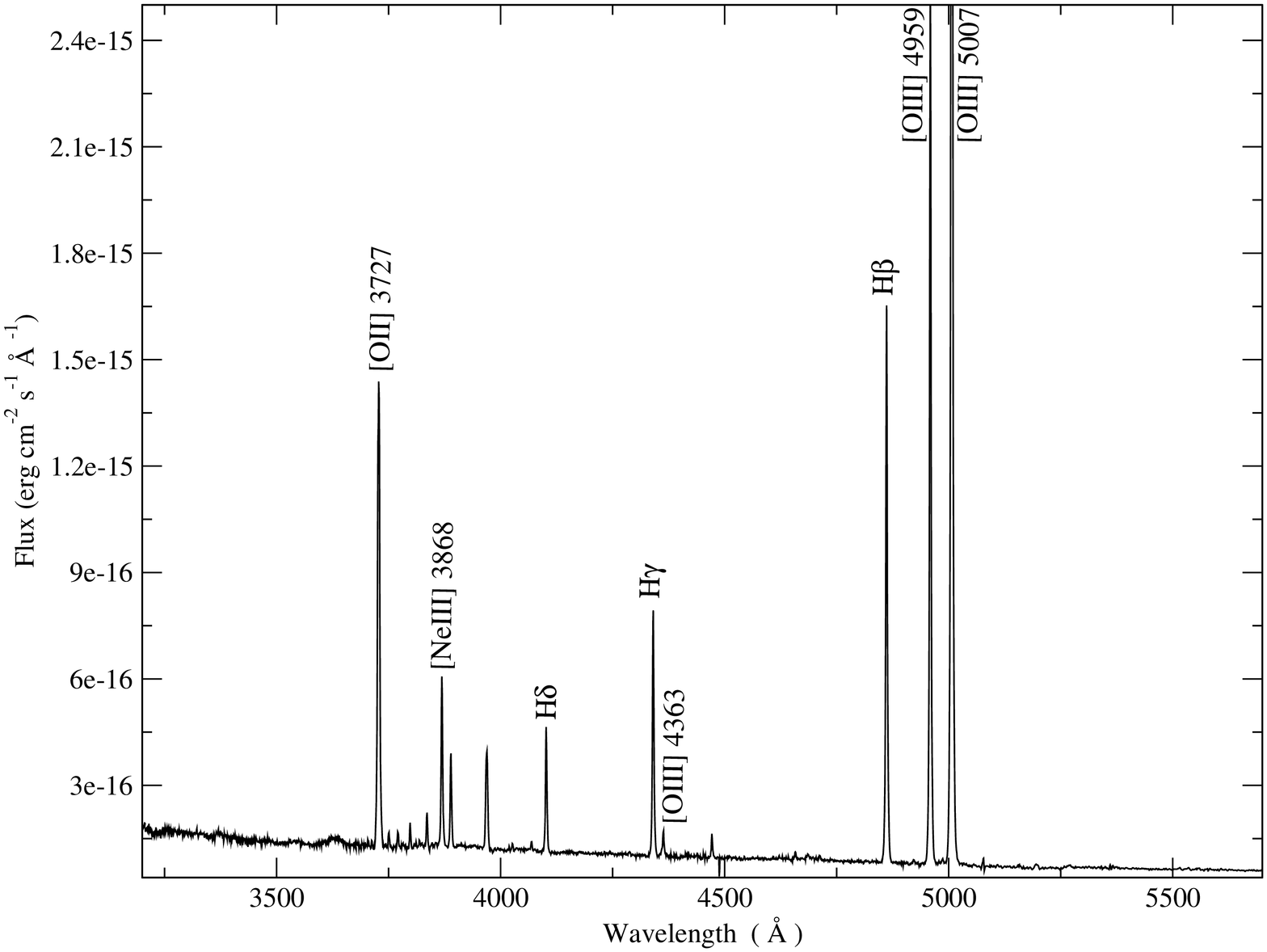}\\
\includegraphics[width=.31\textwidth,height=.63\textwidth,angle=270]{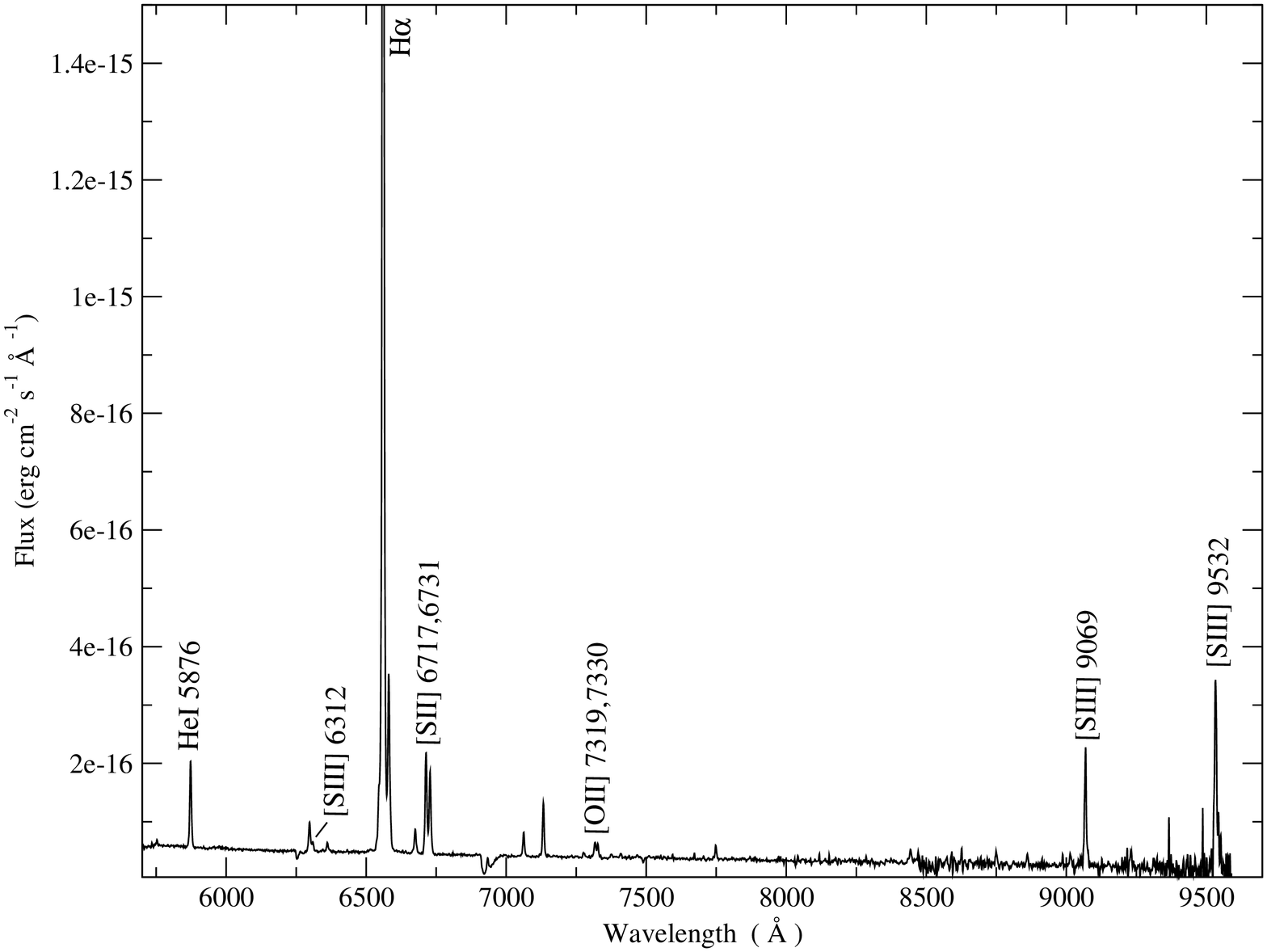}
\caption{WHT spectra of SDSS J002101.03+005248.1 in the rest frame for the two
  observed ranges.}
\label{comp0390}
\end{figure*}

In the last step, the spectra were corrected for atmospheric extinction and flux calibrated. 
For the blue spectra, four standard star observations were used, allowing a good
spectrophotometric calibration with an estimated accuracy of about 5\%. 
Unfortunately, only one standard star could be used for the calibration of
the red spectra. Nevertheless, after flux calibration, in the overlapping 
region of the spectra taken with each arm, the agreement  in the average
continuum level was good.

\begin{figure*}
\includegraphics[width=.31\textwidth,height=.63\textwidth,angle=270]{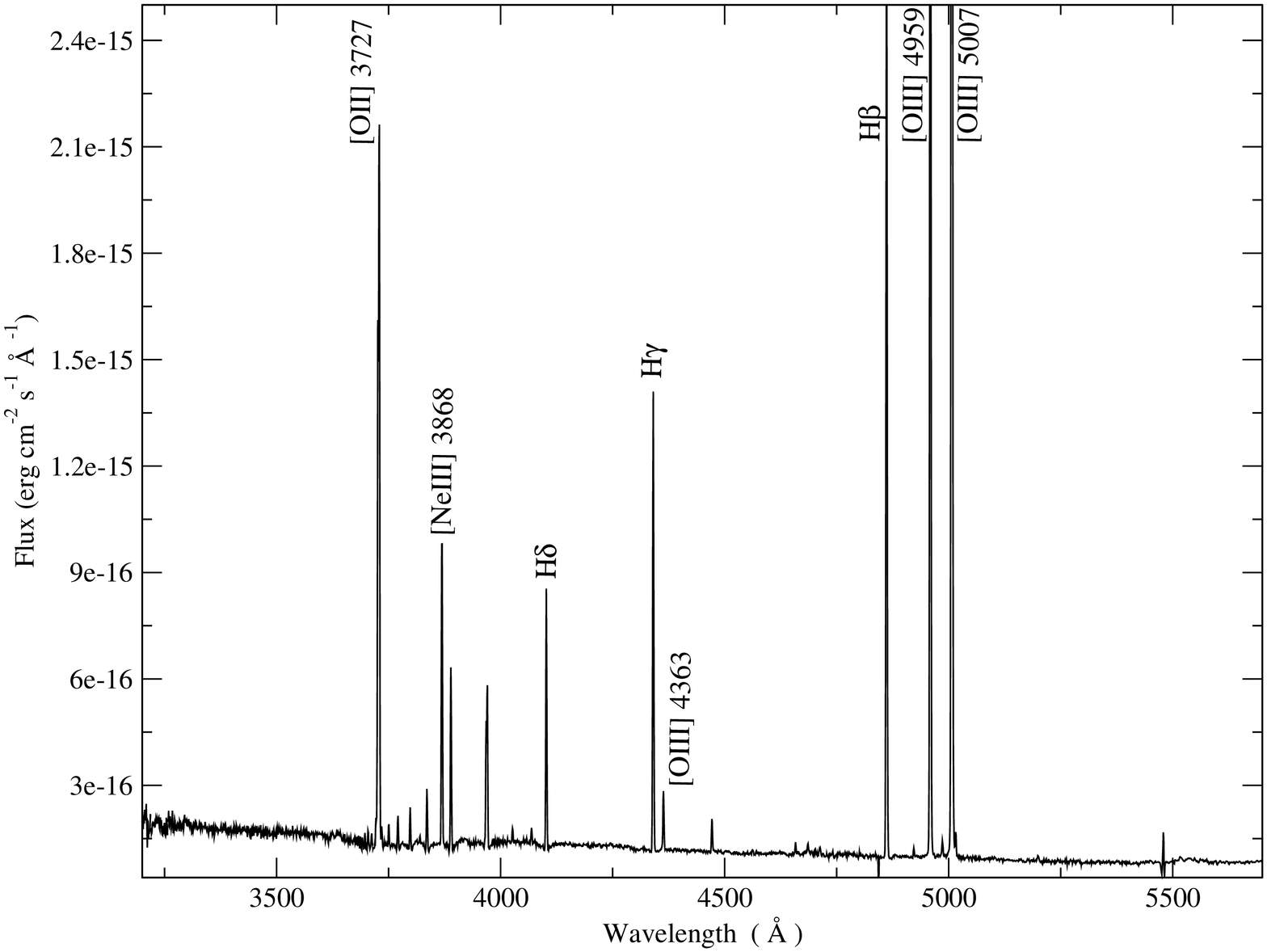}\\
\includegraphics[width=.31\textwidth,height=.63\textwidth,angle=270]{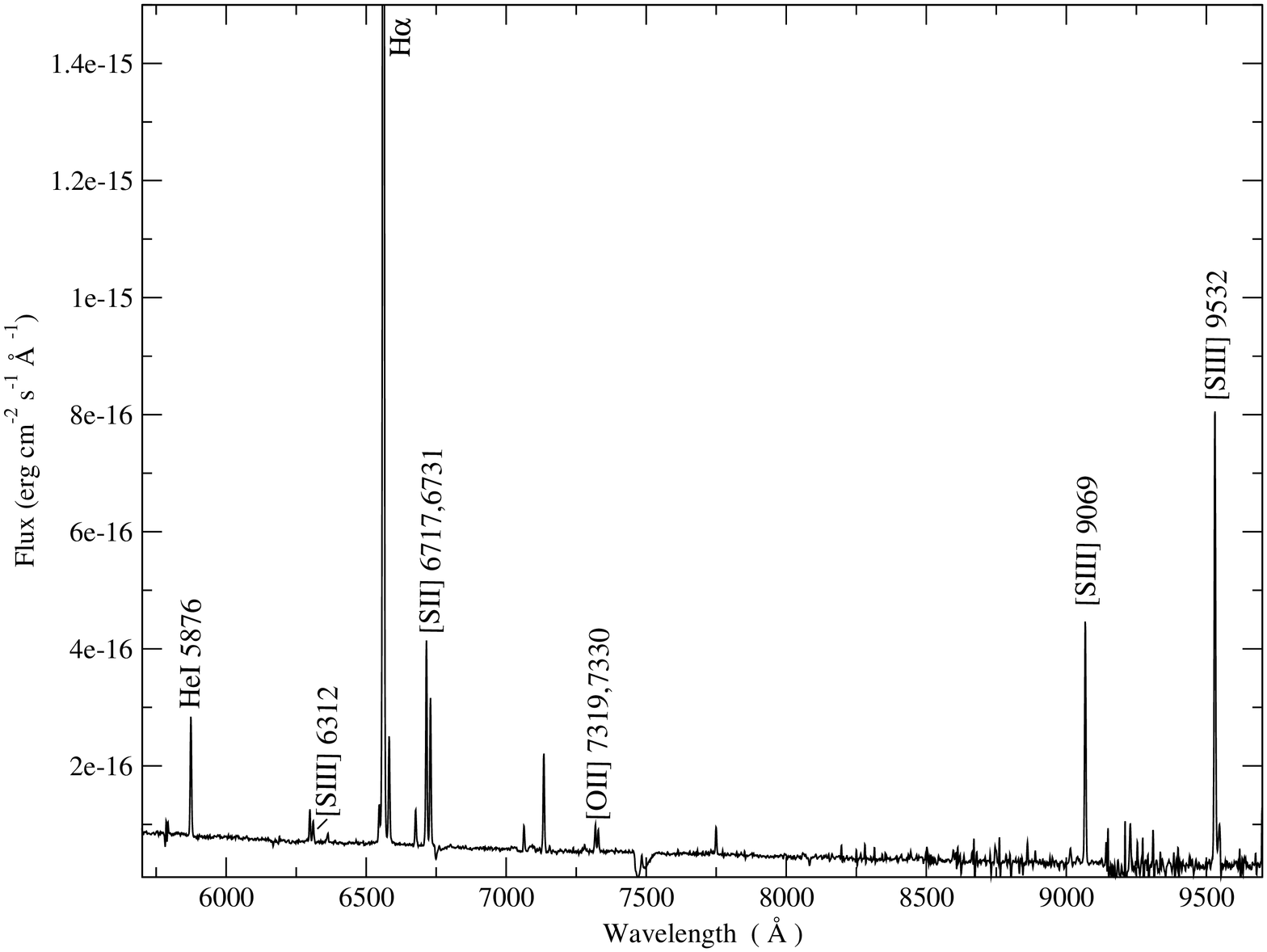}
\caption{WHT spectra of SDSS J003218.60+150014.2 in the rest frame for the two
  observed ranges.}
\label{comp0417}
\end{figure*}
\begin{figure*}
\includegraphics[width=.31\textwidth,height=.63\textwidth,angle=270]{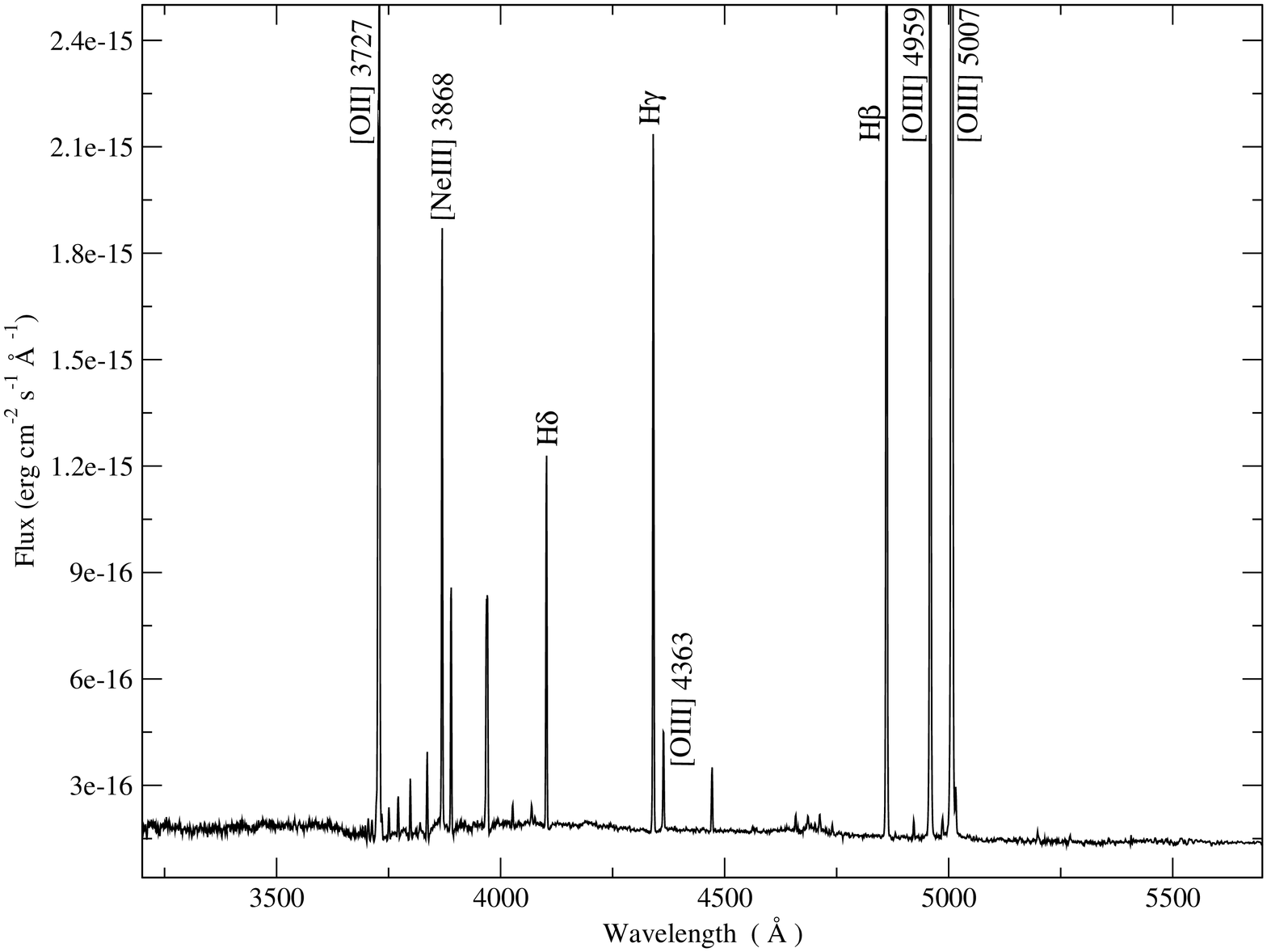}\\
\includegraphics[width=.31\textwidth,height=.63\textwidth,angle=270]{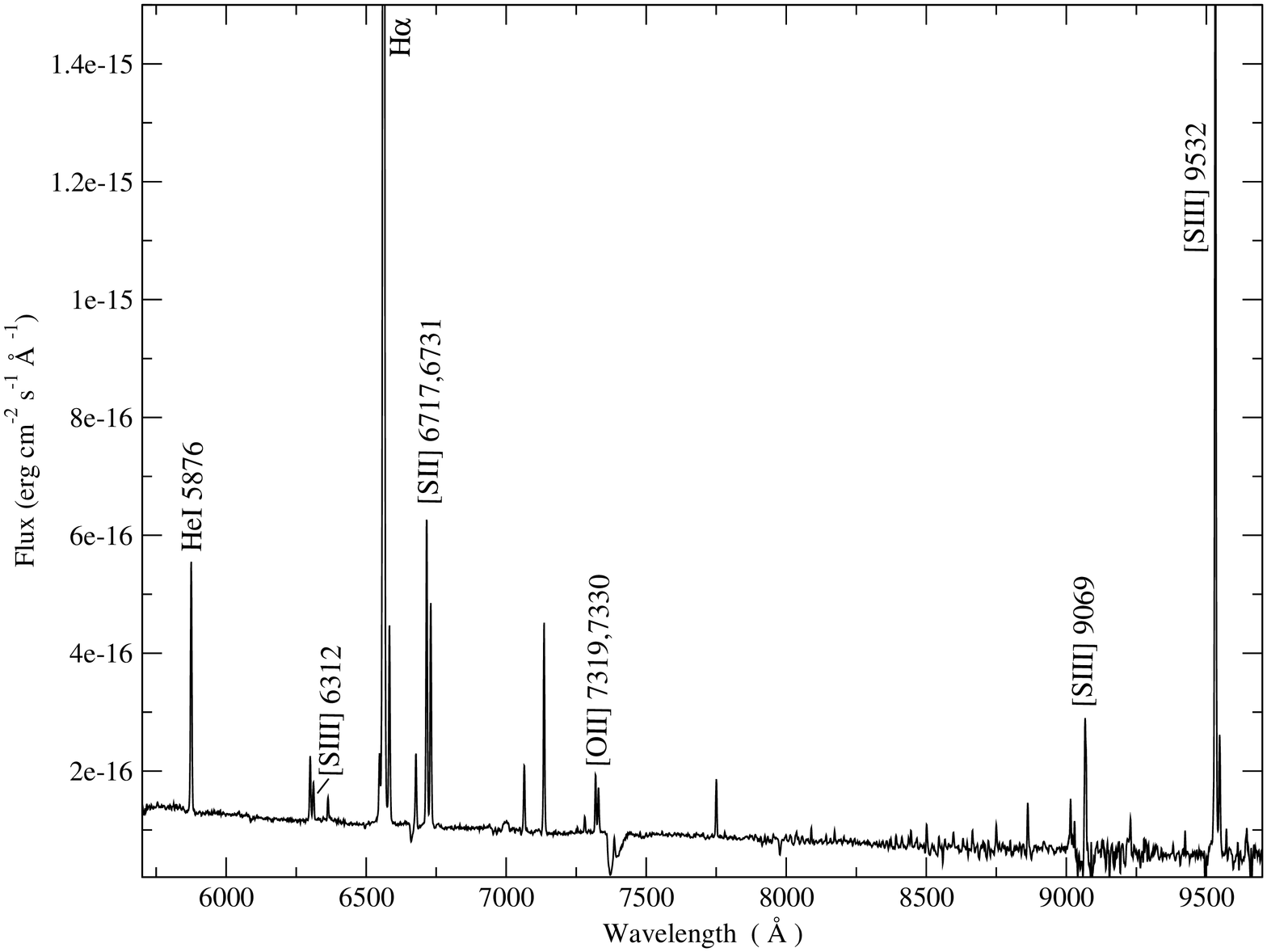}
\caption{WHT spectra of SDSS J162410.11-002202.5 in the rest frame for the two
  observed ranges.} 
\label{comp0364}
\end{figure*}

\section {Results}
\label{measure}

The WHT spectra of the observed galaxies with some of the relevant identified
emission lines are shown in Figs. \ref{comp0390}, \ref{comp0417} and
\ref{comp0364}. Each figure is split into two panels, showing the blue (upper
panel) and red (lower panel) spectral ranges. 

The  emission line fluxes were measured using the SPLOT task of IRAF and are
listed for the three observed galaxies in tables \ref{inten1}, \ref{inten2}, and
\ref{inten3}.  Column 1 of each table shows the wavelength and the name of the
measured lines, as referred in \cite{2004ApJS..153..501G}. The observed
emission line fluxes, $F(\lambda)$ (in  units of  H$\beta$ flux = 1000)
with their corresponding errors, are presented in column 2. The measured
equivalent widths (EW) are listed in column 3.  

We have used two different ways to integrate the flux of
a given line: (1) in the case of an isolated line or two blended and unresolved
lines the intensity was calculated integrating between two points given by the
position of the local continuum placed by eye; (2) if two lines are blended, but
they can be resolved, we have used a multiple gaussian fit procedure to estimate
individual fluxes. Following \cite{1994ApJ...437..239G},
\cite*{2002MNRAS.329..315C} and 
\cite{2003MNRAS.346..105P}, the statistical errors associated with the observed
emission fluxes have been calculated using the expression
$\sigma_{l}$\,=\,$\sigma_{c}$N$^{1/2}$[1 + EW/(N$\Delta$)]$^{1/2}$; where
$\sigma_{l}$ is  
the error in the observed line flux, $\sigma_{c}$ represents the standard
deviation in a box near the measured emission line and stands for the error in
the continuum placement, N is the number of pixels used in the measurement of 
the line flux, EW is the line equivalent width, and $\Delta$ is the wavelength
dispersion in angstroms per pixel. 
There are several lines affected by bad pixels, internal reflections or charge
transfer in the CCD, telluric emission lines or atmospheric absorption
lines. These cause the errors to increase, and, in some cases, they are
impossible to quantify. In these cases we do not include these lines in the tables 
nore in our calculations.  The  only exception is the
emission line [S{\sc iii}]\,$\lambda$\,9069 for SDSS J162410.11-002202.5
which is affected by the strong narrow water-vapor lines present in the
$\lambda$\,9300\,-\,9500 wavelength region \citep*{1985MNRAS.212..737D}.
We have listed the value of the measurement of this line in table
\ref{inten3}, but all the physical parameters depending on its intensity
were calculated using the theoretical ratio between this line and
[S{\sc iii}]\,$\lambda$\,9532, I(9069) $\approx$ 2.44
$\times$ I(9532) \citep{1989agna.book.....O}. 
In some cases there is an observable line (e.g., [Cl{\sc
    iii}]\,$\lambda\lambda$\,5517,5537, several carbon recombination lines,
Balmer or Paschen lines) for which it is impossible to give a precise
measurement. This might be due to a low signal to noise between the line and the
surrounding continuum. This is also the case for the Paschen jump that could not be measured 
even though it was observed, because it was very difficult to locate the
continuum at both sides of the discontinuity with an acceptable accuracy.  

The spectrum of SDSS J002101.03+005248.1 presents very wide 
lines (FWHM\,$\approx$\,7.5\,\AA\ 
for $\lambda$\,$\approx$\,6600\,\AA) for the
expected velocity dispersion in a low mass galaxy of this type. This could be
due to an intrinsic velocity dispersion in this object, the interaction with
another unobservable object or a projection effect on the line of sight. There
are \HII\ galaxies that in fact are multiple systems, with two or more 
components, despite their ``a priory" assumption of compactness
\citep{1966ApJ...143..192Z,1970ApJ...162L.155S}. In some cases, 
these systems show some evidence of interaction among their
components \citep*{1997MNRAS.288...78T}. { For instance, IIZw40, which was
first classified as a compact emission line galaxy by
\cite{1970ApJ...160..405S}, when observed with enough spatial resolution
showed to be the merge of two separate subsystems \citep*{1982MNRAS.198..535B}.}
As a consequence, lines that should be resolved are blended in the
spectrum of SDSS J002101.03+005248.1. Such is the case of H$\alpha$ and 
[N{\sc ii}]\,$\lambda$\,6548\,\AA\
emission lines. We have resorted to the theoretical ratio, I(6584)
$\approx$ 3\,$\times$\,I(6548), to decontaminate the observed flux of H$\alpha$ by
the emission of [N{\sc ii}]\,$\lambda$\,6548 and to derive the electron
temperature of [N{\sc ii}].
 
\begin{figure}
\centering
\includegraphics[width=.37\textwidth,angle=270,clip=]{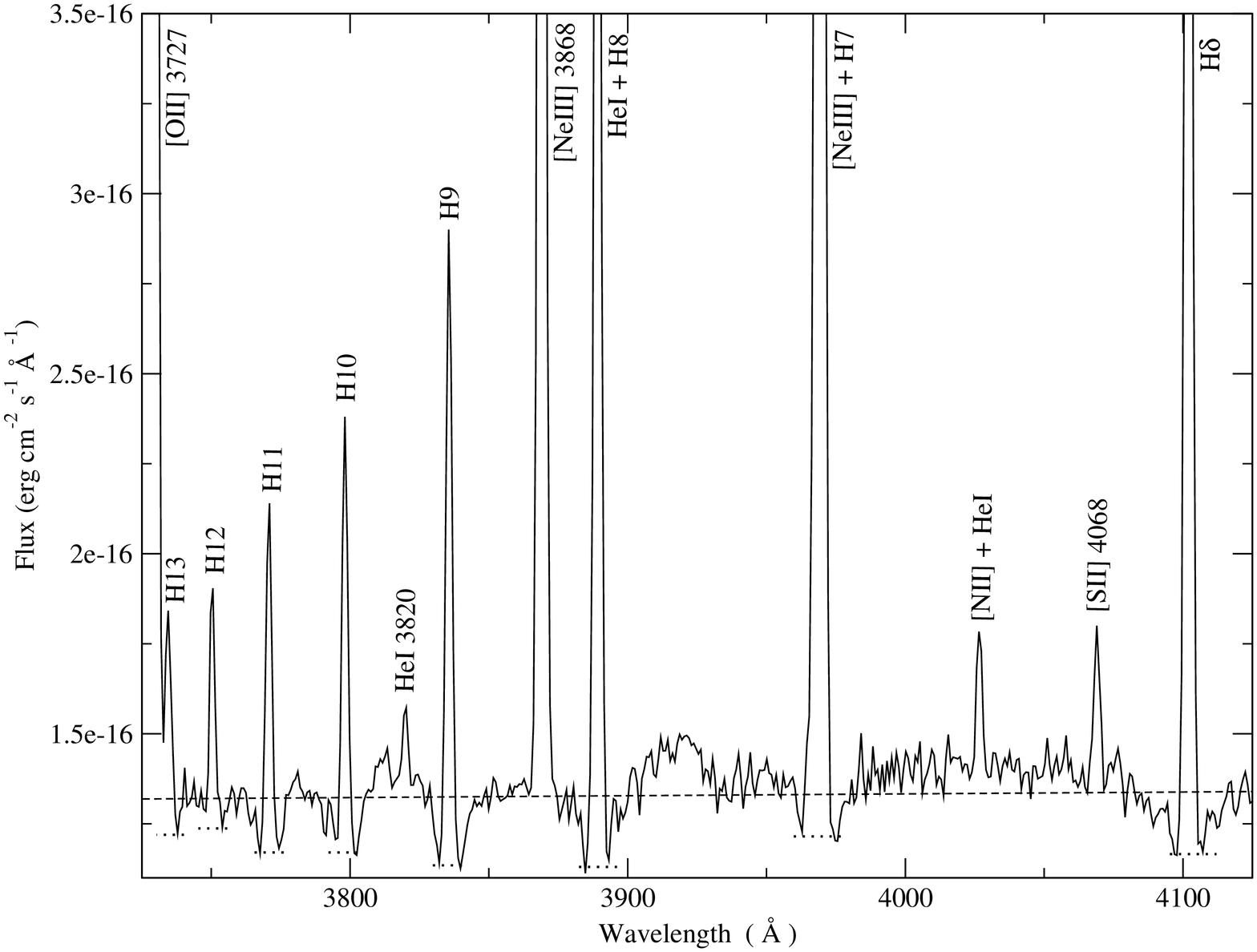}\\
\includegraphics[width=.37\textwidth,angle=270,clip=]{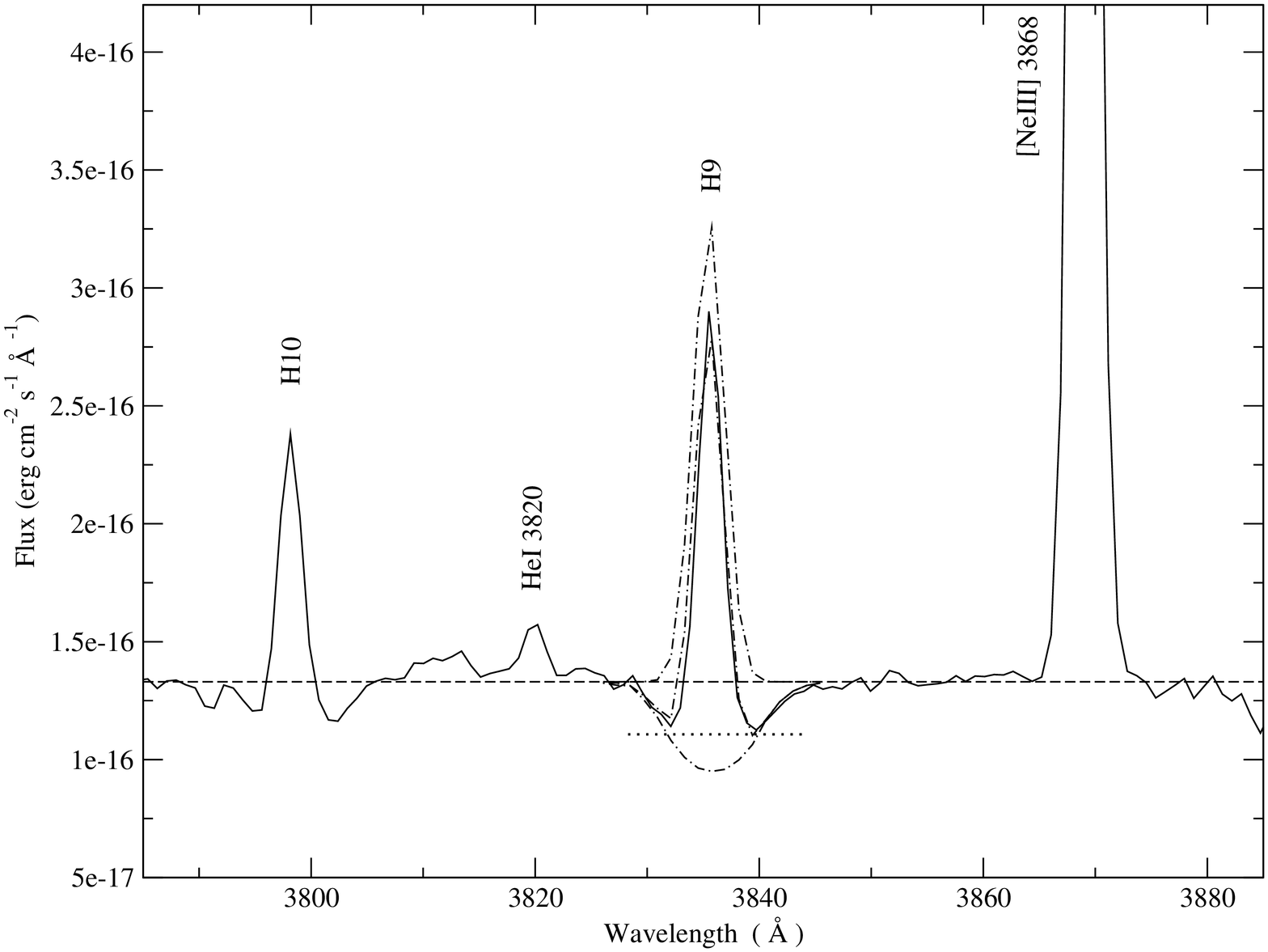}
\caption{{ Upper panel: Section of the spectrum of SDSS J003218.60+150014.2 taken
  with the WHT. The spectrum is in the rest frame and its spectral range is from
  3725 to 4125\,\AA. We can appreciate the presence of absorption features
  originated in an oldish stellar population which mainly affect the Balmer
  emission lines. Lower panel: Section of the same spectrum with a spectral range
  from 3785 to 3885\,\AA. We have superposed the fit to H9 made using the
  ngaussfit task from IRAF (dashed-dotted lines). For both panels: The
  dashed line traces the continuum and the dotted lines show the pseudo-continuum
  used to measure the Balmer emission lines.}}
\label{under}
\end{figure}

A conspicuous underlying stellar population is easily appreciable by the
presence of absorption features that depress the Balmer and Paschen emission
lines. The upper panel of Fig. \ref{under} shows an example of this effect for the
Balmer lines (H13 to H$\delta$) on an enlargement of the spectrum of SDSS 
J003218.60+150014.2, {the object that presents the most relevant and appreciable
absorption lines}. The pseudo-continuum used to measure the line fluxes is
also shown. We can clearly see the wings of the absorption lines implying that,
even though we have used a pseudo-continuum, there is an absorbed fraction of
the emitted flux that we are not able to measure { with an acceptable
accuracy} (see discussion in D\'iaz 1988). This fraction is not the same
for all lines, nore the ratios between the absorbed fractions and the emissions
are the same. { In order to quantify the effect of the underlying absorption
on the measured emission line intensities, we have performed a multi-gaussian
fit to the absorption and emission components seen in this galaxy. The fitting
can be seen in the lower panel of Fig. \ref{under}. The difference between the
measurements of the absorption subtracted lines and the ones obtained with the
use of the pseudo-continuum is, for all Balmer lines, within the observational
errors and, in fact, the additional fractional error introduced by the
subtraction of the absorption component is almost inappreciable for the
stronger lines. In the other two galaxies, the absorption wings in the Balmer
lines are not prominent enough as to provide sensible results by the
multi-gaussian component fitting. Therefore, we have doubled the error derived
using the  the expression for the statistical errors associated with the
observed emission fluxes, ($\sigma_{l}$).} 

The absorption features of the underlying stellar population may
also affect the helium emission lines to some extent. However, these absorption
lines are narrower than those of hydrogen \cite[see, for example,
][]{2005MNRAS.357..945G}, and therefore we cannot see their wings at both sides
of the emission and we cannot define a pseudo-continuum to measure the line fluxes. 

The reddening coefficient ($C$(H$\beta$)) has been calculated assuming the
galactic extinction law of \cite{1972ApJ...172..593M} with
$R_v$=3.2. $C(H\beta)$ was obtained in each case by performing a least square
fit to the ratio between $F(\lambda)$ and $F(H\beta)$ to the theoretical values
computed by \cite{1995MNRAS.272...41S} using an iterative method to estimate
$T_e$ and  $n_e$ in each case. We have taken n$_e$ equal to n([S{\sc ii}]).  Due
to the large error introduced by the presence of the underlying stellar
population (see discussion above) only the four strongest Balmer emission lines
(H$\alpha$, H$\beta$, H$\gamma$ and H$\delta$) have been taken into account. The
values obtained for $C(H\beta)$ and their corresponding errors, considered to be
the uncertainties of the least square fittings, are listed in tables
\ref{inten1}, \ref{inten2}, and \ref{inten3} for each of the observed objects.

The emission line intensities corrected for reddening, relative to H$\beta$
($I(\lambda)$), and their corresponding errors are listed in column  4 of tables
\ref{inten1}, \ref{inten2}, and \ref{inten3}. The errors were obtained
propagating in quadratures the observational errors in the emission line fluxes
and the reddening constant uncertainties. We have not taken into account errors
in the theoretical intensities since they are much lower than the observational
ones. Finally, the values listed in Column 5 of the tables indicate the
fractional error in the line intensities calculated as explained above. 
{ These errors vary from a few percent for the more intense nebular
emission lines (e.g. [O{\sc iii}]\,$\lambda\lambda$\,4959,5007, [S{\sc
    ii}]\,$\lambda\lambda$\,6717,6731 or the strongest Balmer emission lines) to
10-16\,\% for the weakest lines that have less contrast with the continuum noise
(e.g. He{\sc i}\,$\lambda\lambda$\,3820,7281, [Ar{\sc iv}]\,$\lambda$\,4740 or
O{\sc i}\,$\lambda$\,8446). For the auroral lines, the fractional errors are
between $\sim$3 and $\sim$9\,\%.} 

\begin{table*}
{\small
\caption{Relative observed and reddening corrected line intensities
  [$F(H\beta)$=$I(H\beta)$=1000] for  SDSS J002101.03+005248.1} 
\label{inten1}
\begin{center}
\begin{tabular}{|l|cccc|cccc|}
\hline
                           & \multicolumn{8}{c|}{SDSS J002101.03+005248.1 - spSpec-51900-0390-445} \\
\cline{2-9}
\multicolumn{1}{|c|}{$\lambda$  ({\AA})}  & \multicolumn{4}{c|}{WHT} & \multicolumn{4}{c|}{SDSS}   \\
 & $F(\lambda)$  &  -EW(\AA)  & $I(\lambda)$ & Error (\%) & $F(\lambda)$  & -EW(\AA)  & $I(\lambda)$ & Error (\%) \\
\hline

3697  H17                    &    4$\pm$1  &   0.3 &    6$\pm$1  & 11.7 &    ---       &  ---  &     ---      & ---   \\
3704  H16+He{\sc i}          &   11$\pm$1  &   0.7 &   15$\pm$1  &  9.9 &    ---       &  ---  &     ---      & ---   \\
3712  H15                    &    8$\pm$1  &   0.5 &   12$\pm$2  & 15.5 &    ---       &  ---  &     ---      & ---   \\
3727  [O{\sc ii}]$^b$        & 1190$\pm$9  &  71.1 & 1634$\pm$19 &  1.2 &  1598$\pm$15 & 95.0 &  1789$\pm$20 &  1.1  \\
3750  H12                    &   21$\pm$2  &   1.3 &   29$\pm$3  &  9.4 &    27$\pm$3  &  1.8 &    30$\pm$4  & 12.4  \\
3770  H11                    &   28$\pm$2  &   1.7 &   38$\pm$3  &  7.8 &    30$\pm$3  &  2.0 &    33$\pm$4  & 11.3  \\
3798  H10                    &   31$\pm$2  &   2.0 &   42$\pm$3  &  6.7 &    35$\pm$3  &  2.3 &    39$\pm$4  &  9.5  \\
3835  H9                     &   49$\pm$5  &   3.2 &   65$\pm$6  &  9.7 &    57$\pm$6  &  4.0 &    63$\pm$7  & 11.1  \\
3868  [Ne{\sc iii}]          &  294$\pm$6  &  17.7 &  388$\pm$8  &  2.1 &   346$\pm$6  & 21.6 &   382$\pm$7  &  1.9  \\
3889  He{\sc i}+H8           &  155$\pm$9  &   9.7 &  203$\pm$11 &  5.6 &   169$\pm$9  & 11.2 &   186$\pm$10 &  5.5  \\
3968  [Ne{\sc iii}]+H7       &  226$\pm$10 &  15.2 &  290$\pm$13 &  4.4 &   248$\pm$14 & 17.2 &   271$\pm$15 &  5.5  \\
4026  [N{\sc ii}]+He{\sc i}  &   12$\pm$1  &   0.8 &   15$\pm$2  & 10.9 &    10$\pm$1  &  0.6 &    10$\pm$1  & 11.9  \\
4068  [S{\sc ii}]            &   11$\pm$1  &   0.8 &   14$\pm$1  &  8.6 &    12$\pm$1  &  0.8 &    12$\pm$1  &  9.7  \\
4102  H$\delta$              &  213$\pm$6  &  15.5 &  265$\pm$7  &  2.8 &   230$\pm$7  & 17.1 &   249$\pm$8  &  3.2  \\
4340  H$\gamma$              &  423$\pm$7  &  33.6 &  499$\pm$9  &  1.7 &   442$\pm$9  & 36.5 &   469$\pm$9  &  2.0  \\
4363  [O{\sc iii}]           &   47$\pm$2  &   3.8 &   56$\pm$3  &  4.9 &    40$\pm$2  &  3.3 &    42$\pm$2  &  5.2  \\
4471  He{\sc i}              &   37$\pm$3  &   3.0 &   42$\pm$3  &  7.9 &    36$\pm$2  &  3.1 &    38$\pm$3  &  6.9  \\
4658  [Fe{\sc iii}]          &   10$\pm$1  &   0.8 &   10$\pm$1  &  7.1 &    11$\pm$1  &  0.9 &    11$\pm$1  & 12.3  \\
4686  He{\sc ii}             &    6$\pm$1  &   0.5 &    7$\pm$1  &  8.1 &     4$\pm$1  &  0.3 &     4$\pm$1  &  7.0  \\
4713  [Ar{\sc iv}]+He{\sc i} &    6$\pm$1  &   0.6 &    7$\pm$1  &  5.6 &     5$\pm$1  &  0.5 &     5$\pm$1  & 10.5  \\
4740  [ArIV]                 &    5$\pm$1  &   0.4 &    5$\pm$1  & 12.4 &      ---     &  ---  &      ---     &  ---  \\
4861  H$\beta$               & 1000$\pm$9  &  97.0 & 1000$\pm$9  &  0.9 &  1000$\pm$9  &108.1 &  1000$\pm$9  &  0.9  \\
4881  [Fe{\sc iii}]          &     ---     &  ---   &     ---     & ---  &     4$\pm$1  &  0.4 &     4$\pm$1  & 11.2  \\
4921  He{\sc i}              &    6$\pm$1  &   0.6 &    6$\pm$1  &  8.6 &     8$\pm$1  &  0.9 &     8$\pm$1  &  6.9  \\
4959  [O{\sc iii}]           & 1575$\pm$13 & 151.5 & 1532$\pm$13 &  0.8 &  1518$\pm$10 &160.0 &  1503$\pm$10 &  0.6  \\
4986  [Fe{\sc iii}]$^c$      &    9$\pm$1  &   0.9 &    9$\pm$1  & 10.2 &    10$\pm$1  &  1.1 &    10$\pm$1  & 10.5  \\
5007  [O{\sc iii}]           & 4514$\pm$26 & 439.4 & 4334$\pm$26 &  0.6 &  4583$\pm$26 &489.4 &  4516$\pm$26 &  0.6  \\
5199  [N{\sc i}]             &   13$\pm$1  &   1.5 &   12$\pm$1  & 11.2 &    12$\pm$1  &  1.4 &    11$\pm$1  &  8.1  \\
5270  [Fe{\sc iii}]$^a$      &     ---     &  ---   &     ---     &  --- &     9$\pm$1  &  1.1 &     9$\pm$1  &  9.1  \\
5755  [N{\sc ii}]            &    7$\pm$1  &   0.9 &    5$\pm$1  &  7.1 &     7$\pm$1  &  1.0 &     6$\pm$1  &  6.9  \\
5876  He{\sc i}              &  161$\pm$8  &  22.3 &  127$\pm$7  &  5.2 &   137$\pm$7  & 22.0 &   125$\pm$6  &  5.0  \\
6300  [O{\sc i}]             &   54$\pm$2  &   8.2 &   39$\pm$1  &  3.3 &    47$\pm$3  &  8.4 &    42$\pm$3  &  6.1  \\
6312  [S{\sc iii}]           &   17$\pm$1  &   2.6 &   12$\pm$1  &  4.8 &    16$\pm$1  &  2.8 &    14$\pm$1  &  6.8  \\
6364  [O{\sc i}]             &   15$\pm$1  &   2.4 &   11$\pm$1  &  5.2 &    14$\pm$1  &  2.6 &    12$\pm$1  &  6.8  \\
6548  [N{\sc ii}]           &      ---    &   ---  &     ---     &  --- &   ---   &  --- &   ---  & ---  \\
6563  H$\alpha$              & 4160$\pm$39 & 658.3 & 2886$\pm$40 &  1.4 &  3236$\pm$8  &607.1 &  2841$\pm$22 &  0.8  \\
6584  [N{\sc ii}]            &  377$\pm$12 &  59.6 &  260$\pm$9  &  3.3 &   268$\pm$6  & 50.8 &   235$\pm$5  &  2.2  \\
6678  He{\sc i}              &   47$\pm$2  &   7.7 &   32$\pm$1  &  4.6 &    38$\pm$2  &  7.4 &    33$\pm$2  &  6.0  \\
6717  [S{\sc ii}]            &  201$\pm$7  &  33.8 &  136$\pm$5  &  3.4 &   180$\pm$5  & 35.4 &   157$\pm$5  &  3.0  \\
6731  [S{\sc ii}]            &  159$\pm$7  &  27.2 &  107$\pm$5  &  4.3 &   136$\pm$5  & 26.9 &   118$\pm$4  &  3.6  \\
7065  He{\sc i}              &   39$\pm$2  &   7.2 &   25$\pm$1  &  4.8 &    30$\pm$2  &  6.5 &    25$\pm$1  &  5.8  \\
7136  [Ar{\sc iii}]          &   97$\pm$4  &  18.5 &   62$\pm$2  &  3.9 &    74$\pm$3  & 17.5 &    63$\pm$2  &  4.0  \\
7155  [Fe{\sc ii}]           &      ---    &   ---  &     ---     &  --- &     3$\pm$1  &  0.8 &     3$\pm$1  & 12.5  \\
7254  O{\sc i}               &      ---    &   ---  &     ---     &  --- &     3$\pm$1  &  0.8 &     3$\pm$1  & 13.1  \\
7281  He{\sc i}$^a$          &   10$\pm$1  &   1.9 &    6$\pm$1  &  7.0 &     8$\pm$1  &  2.1 &     7$\pm$1  &  5.6  \\
7319  [O{\sc ii}]$^d$        &   31$\pm$1  &   6.2 &   20$\pm$1  &  3.2 &    24$\pm$1  &  5.8 &    21$\pm$1  &  2.5  \\
7330  [O{\sc ii}]$^e$        &   21$\pm$1  &   4.2 &   13$\pm$1  &  4.5 &    18$\pm$1  &  4.2 &    15$\pm$1  &  3.6  \\
7378  [Ni{\sc ii}]           &    6$\pm$1  &   1.2 &    4$\pm$1  &  8.4 &     4$\pm$1  &  0.9 &     3$\pm$1  &  8.9  \\
7412  [Ni{\sc ii}]           &    7$\pm$1  &   1.4 &    4$\pm$1  &  7.9 &      ---     & ---   &     ---      &  ---  \\
7751  [Ar{\sc iii}]          &   23$\pm$2  &   5.1 &   14$\pm$1  &  6.9 &    22$\pm$2  &  6.1 &    19$\pm$1  &  7.4  \\
8446  O{\sc i}               &   28$\pm$2  &   7.2 &   16$\pm$1  &  7.6 &      ---     &  ---  &      ---     &  ---  \\
9069  [S{\sc iii}]           &  230$\pm$11 &  71.8 &  119$\pm$6  &  5.0 &      ---     &  ---  &      ---     &  ---  \\
9532  [S{\sc iii}]           &  476$\pm$16 & 504.4 &  239$\pm$9  &  3.8 &      ---     &  ---  &      ---     &  ---  \\

\hline
I(H$\beta$)(erg\,seg$^{-1}$\,cm$^{-2}$) & \multicolumn{4}{c|}{2.47\,$\times$\,10$^{-14}$} &  \multicolumn{4}{c|}{3.38\,$\times$\,10$^{-14}$} \\
C(H$\beta$) & \multicolumn{4}{c|}{0.51$\pm$0.01} &  \multicolumn{4}{c|}{0.18$\pm$0.01} \\
\hline

\end{tabular}
\end{center}

\medskip
\begin{flushleft} 
$^a$\,possibly blend with an unknown line; $^b$\,[O{\sc ii}]
$\lambda\lambda$\,3726\,+\,3729;  $^c$\,[Fe{\sc iii}]
$\lambda\lambda$\,4986\,+\,4987; $^d$\,[O{\sc ii}]
$\lambda\lambda$\,7318\,+\,7320; $^e$\,[O{\sc ii}]
$\lambda\lambda$\,7330\,+\,7331.
\end{flushleft} 
}
\end{table*}

\begin{table*}
{\small
\caption{Relative observed and reddening corrected  emission line intensities [$F(H\beta)$=$I(H\beta)$=1000] for SDSS J003218.60+150014.2.}
\label{inten2}
\begin{center}
\begin{tabular}{|l|cccc|cccc|}
\hline
                           & \multicolumn{8}{c|}{SDSS J003218.60+150014.2 - spSpec-51817-0418-302} \\
\cline{2-9}
\multicolumn{1}{|c|}{$\lambda$  ({\AA})}  & \multicolumn{4}{c|}{WHT} & \multicolumn{4}{c|}{SDSS}   \\
 & $F(\lambda)$  &  -EW(\AA)  & $I(\lambda)$ & Error (\%) & $F(\lambda)$  & -EW(\AA)  & $I(\lambda)$ & Error (\%) \\
\hline

3687 H19                    &   10$\pm$1  &   0.6 &    10$\pm$1  &   8.5 &     ---     &    --- &    ---      &  ---  \\
3697 H17                    &   13$\pm$1  &   0.9 &    14$\pm$1  &   8.6 &     ---     &    --- &    ---      &  ---  \\
3704 H16+He{\sc i}          &   19$\pm$2  &   1.2 &    20$\pm$2  &   8.4 &     ---     &    --- &    ---      &  ---  \\
3712 H15                    &   15$\pm$1  &   0.9 &    16$\pm$1  &   7.5 &     ---     &    --- &    ---      &  ---  \\
3727 [O{\sc ii}]$^b$        & 1466$\pm$7  &  80.0 &  1573$\pm$14 &   0.9 &     ---     &    --- &    ---      &  ---  \\
3734 H13                    &   18$\pm$1  &   1.2 &    20$\pm$1  &   7.5 &     ---     &    --- &    ---      &  ---  \\
3750 H12                    &   20$\pm$2  &   1.3 &    21$\pm$2  &   7.7 &     ---     &    --- &    ---      &  ---  \\  
3770 H11                    &   34$\pm$2  &   2.3 &    37$\pm$2  &   4.6 &   45$\pm$3  &   3.0  &   51$\pm$4  &  7.3  \\
3798 H10                    &   42$\pm$2  &   2.8 &    45$\pm$3  &   5.7 &   52$\pm$4  &   3.6  &   59$\pm$4  &  6.8  \\
3820 He{\sc i}              &    7$\pm$1  &   0.4 &     7$\pm$1  &  14.3 &     ---     &    --- &    ---      &  ---  \\
3835 H9                     &   61$\pm$5  &   4.4 &    65$\pm$5  &   8.3 &   59$\pm$5  &   3.7  &   67$\pm$5  &  7.7  \\
3868 [Ne{\sc iii}]          &  350$\pm$9  &  19.9 &   372$\pm$10 &   2.6 &  370$\pm$11 &  19.7  &  418$\pm$12 &  2.9  \\
3889 He{\sc i}+H8           &  184$\pm$12 &  12.5 &   196$\pm$12 &   6.3 &  181$\pm$11 &  10.7  &  204$\pm$13 &  6.2  \\
3968 [Ne{\sc iii}]+H7       &  243$\pm$12 &  17.0 &   257$\pm$13 &   5.0 &  244$\pm$15 &  17.4  &  272$\pm$17 &  6.1  \\
4026 [N{\sc ii}]+He{\sc i}  &   14$\pm$1  &   0.8 &    14$\pm$1  &   7.1 &   15$\pm$2  &   0.9  &   17$\pm$2  & 12.1  \\
4068 [S{\sc ii}]            &   18$\pm$1  &   1.1 &    19$\pm$1  &   5.5 &   13$\pm$1  &   0.8  &   15$\pm$1  &  9.1  \\
4102 H$\delta$              &  236$\pm$9  &  17.2 &   247$\pm$9  &   3.7 &  237$\pm$8  &  17.5  &  261$\pm$9  &  3.3  \\
4340 H$\gamma$              &  482$\pm$10 &  33.8 &   500$\pm$10 &   2.0 &  440$\pm$8  &  32.5  &  473$\pm$9  &  1.9  \\
4363 [O{\sc iii}]           &   60$\pm$3  &   4.2 &    62$\pm$3  &   4.8 &   60$\pm$4  &   4.2  &   65$\pm$4  &  6.4  \\
4471 He{\sc i}              &   36$\pm$2  &   2.7 &    37$\pm$2  &   5.6 &   34$\pm$3  &   2.4  &   36$\pm$3  &  8.5  \\
4658 [Fe{\sc iii}]          &    9$\pm$1  &   0.7 &     9$\pm$1  &   9.1 &   10$\pm$1  &   0.7  &   10$\pm$1  &  8.1  \\
4686 He{\sc ii}             &   13$\pm$1  &   1.0 &    13$\pm$1  &   6.3 &   14$\pm$1  &   1.0  &   14$\pm$1  &  8.6  \\
4713 [Ar{\sc iv}]+He{\sc i} &   11$\pm$1  &   0.9 &    11$\pm$1  &   5.5 &   11$\pm$1  &   0.8  &   11$\pm$1  & 11.4  \\
4740 [Ar{\sc iv}]           &    5$\pm$1  &   0.4 &     5$\pm$1  &  12.9 &    3$\pm$1  &   0.3  &    3$\pm$1  & 11.9  \\
4861 H$\beta$               & 1000$\pm$8  &  89.6 &  1000$\pm$8  &   0.8 & 1000$\pm$9  &  86.7  & 1000$\pm$9  &  0.9  \\
4921 He{\sc i}              &    9$\pm$1  &   0.8 &     9$\pm$1  &   6.5 &   11$\pm$1  &   0.9  &   11$\pm$1  &  9.6  \\
4959 [O{\sc iii}]           & 1592$\pm$13 & 140.7 &  1582$\pm$13 &   0.8 & 1651$\pm$11 & 138.6  & 1631$\pm$11 &  0.6  \\
4986 [Fe{\sc iii}]$^c$      &   16$\pm$1  &   1.4 &    15$\pm$1  &   9.1 &   18$\pm$2  &   1.5  &   18$\pm$2  & 9.2   \\
5007 [O{\sc iii}]           & 4649$\pm$24 & 416.4 &  4607$\pm$24 &   0.5 & 4863$\pm$16 & 412.8  & 4778$\pm$16 &  0.3  \\
5015 He{\sc i}              &   23$\pm$2  &   2.0 &    22$\pm$2  &   8.8 &   24$\pm$2  &   2.1  &   24$\pm$2  &  7.8  \\
5199 [N{\sc i}]             &    9$\pm$1  &   0.9 &     9$\pm$1  &   5.4 &   10$\pm$1  &   0.9  &    9$\pm$1  &  7.7  \\
5270 [Fe{\sc iii}]$^a$      &     ---     &  ---  &      ---     &   --- &    7$\pm$1  &   0.7  &    7$\pm$1  &  8.6  \\
5755 [N{\sc ii}]            &     ---     &  ---  &      ---     &   --- &     ---     &    --- &    ---      &  ---  \\
5876 He{\sc i}              &  131$\pm$6  &  16.0 &   124$\pm$6  &   4.5 &  125$\pm$6  &  15.0  &  113$\pm$5  &  4.4  \\
6300 [O{\sc i}]             &   30$\pm$1  &   4.1 &    28$\pm$1  &   3.9 &   30$\pm$1  &   3.9  &   26$\pm$1  &  2.5  \\
6312 [S{\sc iii}]           &   25$\pm$1  &   3.3 &    23$\pm$1  &   3.7 &   20$\pm$1  &   2.6  &   17$\pm$1  &  3.2  \\
6364 [O{\sc i}]             &   11$\pm$1  &   1.5 &    10$\pm$1  &   5.9 &   10$\pm$1  &   1.3  &    9$\pm$1  &  3.0  \\
6548 [N{\sc ii}]            &   41$\pm$3  &   5.8 &    37$\pm$2  &   6.5 &   37$\pm$2  &   5.2  &   31$\pm$2  &  5.9  \\
6563 H$\alpha$              & 3089$\pm$19 & 435.5 &  2848$\pm$30 &   1.0 & 3309$\pm$16 & 462.8  & 2825$\pm$27 &  0.9  \\
6584 [N{\sc ii}]            &  121$\pm$5  &  17.3 &   112$\pm$5  &   4.4 &  108$\pm$3  &  15.1  &   92$\pm$3  &  3.3  \\
6678 He{\sc i}              &   35$\pm$2  &   5.2 &    33$\pm$2  &   5.4 &   34$\pm$1  &   4.9  &   29$\pm$1  &  2.4  \\
6717 [S{\sc ii}]            &  188$\pm$5  &  30.3 &   173$\pm$5  &   2.9 &  185$\pm$4  &  26.9  &  156$\pm$3  &  2.1  \\
6731 [S{\sc ii}]            &  136$\pm$3  &  21.9 &   125$\pm$3  &   2.5 &  135$\pm$3  &  19.7  &  114$\pm$2  &  2.1  \\
7065 He{\sc i}              &   24$\pm$2  &   4.2 &    22$\pm$2  &   9.6 &   24$\pm$2  &   3.9  &   20$\pm$1  &  6.6  \\
7136 [Ar{\sc iii}]          &  102$\pm$3  &  17.7 &    92$\pm$3  &   3.5 &   91$\pm$4  &  15.0  &   75$\pm$4  &  4.9  \\
7155 [Fe{\sc ii}]           &    5$\pm$1  &   0.9 &     4$\pm$1  &  13.6 &     ---     &    --- &    ---      &  ---  \\
7281 He{\sc i}$^a$          &    6$\pm$1  &   1.1 &     6$\pm$1  &  12.0 &    5$\pm$1  &   0.9  &    4$\pm$1  &  6.3  \\
7319 [O{\sc ii}]$^d$        &   27$\pm$1  &   5.0 &    25$\pm$1  &   4.0 &   24$\pm$1  &   4.1  &   20$\pm$1  &  2.9  \\
7330 [O{\sc ii}]$^e$        &   21$\pm$1  &   3.9 &    19$\pm$1  &   4.6 &   18$\pm$1  &   3.1  &   15$\pm$1  &  3.0  \\
7751 [Ar{\sc iii}]          &   26$\pm$3  &   5.1 &    23$\pm$2  &  10.1 &   21$\pm$1  &   3.9  &   17$\pm$1  &  6.1  \\
8446 O{\sc i}               &    7$\pm$1  &   1.6 &     6$\pm$1  &  15.7 &    6$\pm$1  &   1.5  &    5$\pm$1  & 11.0  \\
8468 P17                    &    3$\pm$1  &   0.8 &     3$\pm$1  &  12.5 &     ---     &    --- &    ---      &  ---  \\
8546 P15                    &    ---     &   --- &      ---     &       &    7$\pm$1  &   1.8  &    5$\pm$1  & 10.8  \\
8599 P14                    &   14$\pm$1  &   3.7 &    12$\pm$1  &  10.7 &   10$\pm$1  &   2.8  &    8$\pm$1  &  8.6  \\
8865 P11                    &   17$\pm$2  &   5.1 &    15$\pm$2  &  11.2 &   16$\pm$2  &   4.6  &   12$\pm$1  &  9.4  \\
9014 P10                    &   25$\pm$3  &   8.1 &    21$\pm$2  &  10.7 &     ---     &    --- &    ---      &  ---  \\
9069 [S{\sc iii}]           &  251$\pm$10 &  76.4 &   217$\pm$10 &   4.4 &     ---     &    --- &    ---      &  ---  \\
9532 [S{\sc iii}]           &  518$\pm$43 & 175.5 &   445$\pm$38 &   8.5 &     ---     &    --- &    ---      &  ---  \\
9547 P8                     &   75$\pm$6  &  43.6 &    64$\pm$5  &   8.5 &     ---     &    --- &    ---      &  ---  \\
 
\hline
I(H$\beta$)(erg\,seg$^{-1}$\,cm$^{-2}$) & \multicolumn{4}{c|}{1.23\,$\times$\,10$^{-14}$} &  \multicolumn{4}{c|}{3.17\,$\times$\,10$^{-14}$} \\
C(H$\beta$) & \multicolumn{4}{c|}{0.11$\pm$0.01} &  \multicolumn{4}{c|}{0.22$\pm$0.01} \\
\hline

\end{tabular}
\end{center}

\medskip
\begin{flushleft} 
$^a$\,possibly blend with an unknown line; $^b$\,[O{\sc ii}]
$\lambda\lambda$\,3726\,+\,3729; $^c$\,[Fe{\sc iii}]
$\lambda\lambda$\,4986\,+\,4987; $^d$\,[O{\sc ii}]
$\lambda\lambda$\,7318\,+\,7320; $^e$\,[O{\sc ii}]
$\lambda\lambda$\,7330\,+\,7331.
\end{flushleft} 
}
\end{table*}

\begin{table*}
{\small
\caption{Relative observed and reddening corrected emission line intensities
  [$F(H\beta)$=$I(H\beta)$=1000] for SDSS J162410.11-002202.5.} 
\label{inten3}
\begin{center}
\begin{tabular}{|l|cccc|cccc|}
\hline
                           & \multicolumn{8}{c|}{SDSS J162410.11-002202.5 - spSpec-52000-0364-187} \\
\cline{2-9}
\multicolumn{1}{|c|}{$\lambda$  ({\AA})}  & \multicolumn{4}{c|}{WHT} & \multicolumn{4}{c|}{SDSS}   \\
 & $F(\lambda)$  &  -EW(\AA)  & $I(\lambda)$ & Error (\%) & $F(\lambda)$  & -EW(\AA)  & $I(\lambda)$ & Error (\%) \\
\hline

3687  H19                    &    4$\pm$1   &    0.3 &     5$\pm$1  &  10.9 &     ---     &   ---  &    ---      & ---   \\
3692  H18                    &    5$\pm$1   &    0.4 &     6$\pm$1  &   9.7 &     ---     &   ---  &    ---      & ---   \\
3697  H17                    &    8$\pm$1   &    0.8 &    11$\pm$1  &   9.2 &     ---     &   ---  &    ---      & ---   \\
3704  H16+He{\sc i}          &   16$\pm$1   &    1.5 &    22$\pm$2  &   7.6 &     ---     &   ---  &    ---      & ---   \\
3712  H15                    &   16$\pm$1   &    1.5 &    22$\pm$2  &   7.8 &     ---     &   ---  &    ---      & ---   \\
3727  [O{\sc ii}]$^b$        & 1066$\pm$13  &   91.0 &  1471$\pm$22 &   1.5 & 1160$\pm$13 &  96.81 & 1480$\pm$19 &  1.3  \\
3750  H12                    &   20$\pm$1   &    1.9 &    28$\pm$2  &   6.7 &   18$\pm$2  &   1.57 &   23$\pm$ 2 &  9.3  \\
3770  H11                    &   24$\pm$2   &    2.2 &    33$\pm$2  &   6.7 &   24$\pm$2  &   2.02 &   30$\pm$ 3 &  8.3  \\
3798  H10                    &   35$\pm$2   &    3.2 &    48$\pm$3  &   6.7 &   36$\pm$3  &   2.93 &   45$\pm$ 4 &  8.4  \\
3820  He{\sc i}              &    8$\pm$1   &    0.7 &    11$\pm$1  &  10.9 &    8$\pm$1  &   0.61 &   10$\pm$ 1 & 10.1  \\
3835  H9                     &   51$\pm$4   &    4.6 &    69$\pm$5  &   7.7 &   49$\pm$3  &   4.17 &   61$\pm$ 4 &  6.6  \\
3868  [Ne{\sc iii}]          &  322$\pm$8   &   25.8 &   428$\pm$11 &   2.6 &  359$\pm$8  &  27.28 &  444$\pm$10 &  2.3  \\
3889  He{\sc i}+H8           &  141$\pm$9   &   12.8 &   186$\pm$11 &   6.1 &  145$\pm$9  &  12.61 &  178$\pm$11 &  6.3  \\
3968  [Ne{\sc iii}]+H7       &  249$\pm$14  &   21.1 &   321$\pm$18 &   5.7 &  238$\pm$13 &  20.46 &  289$\pm$15 &  5.3  \\
4026  [N{\sc ii}]+He{\sc i}  &   13$\pm$1   &    1.0 &    16$\pm$1  &   5.9 &   13$\pm$1  &   1.11 &   16$\pm$ 1 &  6.0  \\
4068  [S{\sc ii}]            &   12$\pm$1   &    0.9 &    15$\pm$1  &   8.2 &   11$\pm$1  &   0.96 &   14$\pm$ 1 & 10.8  \\
4102  H$\delta$              &  207$\pm$6   &   17.0 &   258$\pm$8  &   3.0 &  212$\pm$6  &  19.74 &  251$\pm$ 7 &  2.9  \\
4340  H$\gamma$              &  386$\pm$7   &   34.6 &   457$\pm$8  &   1.8 &  407$\pm$6  &  39.78 &  463$\pm$ 7 &  1.5  \\
4363  [O{\sc iii}]           &   59$\pm$2   &    5.0 &    70$\pm$2  &   3.4 &   53$\pm$1  &   5.02 &   60$\pm$ 1 &  2.3  \\
4471  He{\sc i}              &   37$\pm$1   &    3.1 &    42$\pm$2  &   3.8 &   35$\pm$1  &   3.41 &   39$\pm$ 1 &  3.2  \\
4658  [Fe{\sc iii}]          &    9$\pm$1   &    0.8 &     9$\pm$1  &   8.2 &   10$\pm$1  &   0.96 &   10$\pm$ 1 & 10.0  \\
4686  He{\sc ii}             &    9$\pm$1   &    0.8 &    10$\pm$1  &   7.0 &    8$\pm$1  &   0.75 &    8$\pm$ 1 &  6.7  \\
4713  [Ar{\sc iv}]+He{\sc i} &   11$\pm$1   &    2.0 &    11$\pm$1  &   8.4 &   10$\pm$1  &   1.01 &   10$\pm$ 1 &  7.8  \\
4740  [Ar{\sc iv}]           &    5$\pm$1   &    0.4 &     5$\pm$1  &   8.4 &    5$\pm$1  &   0.48 &    5$\pm$ 0 & 10.5  \\
4861  H$\beta$               & 1000$\pm$11  &  100.6 &  1000$\pm$11 &   1.1 & 1000$\pm$9  & 117.31 & 1000$\pm$ 9 &  0.9  \\
4881  [Fe{\sc iii}]          &    3$\pm$1   &    0.3 &     3$\pm$1  &  10.3 &    9$\pm$1  &   0.31 &    9$\pm$ 1 & 10.6  \\
4921  He{\sc i}              &   10$\pm$1   &    1.0 &    10$\pm$1  &   4.3 &    8$\pm$1  &   0.85 &    8$\pm$ 1 &  8.7  \\
4959  [O{\sc iii}]           & 1993$\pm$14  &  191.7 &  1938$\pm$14 &   0.7 & 2005$\pm$12 & 220.10 & 1963$\pm$12 &  0.6  \\
4986  [Fe{\sc iii}]$^c$      &   13$\pm$1   &    1.3 &    13$\pm$1  &   6.6 &   11$\pm$1  &   1.18 &   10$\pm$ 1 &  6.3  \\
5007  [O{\sc iii}]           & 5882$\pm$28  &  565.3 &  5642$\pm$28 &   0.5 & 6042$\pm$18 & 659.56 & 5855$\pm$18 &  0.3  \\
5015  He{\sc i}              &   29$\pm$3   &    2.8 &    28$\pm$3  &  11.4 &   19$\pm$2  &   2.07 &   18$\pm$ 2 & 10.6  \\
5159  [Fe{\sc ii}]$^a$       &    4$\pm$1   &    0.4 &     4$\pm$1  &  12.6 &     ---     &   ---  &    ---      & ---   \\
5199  [N{\sc i}]             &    8$\pm$1   &    0.9 &     7$\pm$1  &   8.0 &    8$\pm$1  &   0.85 &    7$\pm$ 1 &  8.2  \\
5270  [Fe{\sc iii}]$^a$      &    5$\pm$1   &    0.6 &     5$\pm$1  &  10.2 &    4$\pm$1  &   0.47 &    4$\pm$ 0 &  7.9  \\
5755  [N{\sc ii}]            &    3$\pm$1   &    0.4 &     3$\pm$1  &   8.4 &     ---     &   ---  &    ---      & ---   \\
5876  He{\sc i}              &  172$\pm$4   &   20.4 &   135$\pm$3  &   2.4 &  137$\pm$3  &  21.54 &  113$\pm$ 3 &  2.4  \\
6300  [O{\sc i}]             &   42$\pm$1   &    5.5 &    30$\pm$1  &   2.2 &   36$\pm$1  &   6.41 &   28$\pm$ 1 &  2.2  \\
6312  [S{\sc iii}]           &   25$\pm$1   &    3.3 &    18$\pm$1  &   2.7 &   19$\pm$1  &   3.31 &   15$\pm$ 0 &  3.1  \\
6364  [O{\sc i}]             &   15$\pm$1   &    2.0 &    11$\pm$1  &   3.0 &   13$\pm$1  &   2.24 &   10$\pm$ 0 &  3.3  \\
6548  [N{\sc ii}]            &   49$\pm$2   &    6.8 &    34$\pm$2  &   5.1 &   38$\pm$2  &   6.82 &   28$\pm$ 1 &  4.1  \\
6563  H$\alpha$              & 4053$\pm$33  &  566.3 &  2795$\pm$36 &   1.3 & 3737$\pm$17 & 680.37 & 2820$\pm$24 &  0.9  \\
6584  [N{\sc ii}]            &  135$\pm$3   &   18.8 &    93$\pm$2  &   2.6 &  106$\pm$2  &  19.27 &   80$\pm$ 1 &  1.8  \\
6678  He{\sc i}              &   51$\pm$3   &    7.7 &    35$\pm$2  &   5.9 &   45$\pm$3  &   8.51 &   34$\pm$ 2 &  5.9  \\
6717  [S{\sc ii}]            &  203$\pm$5   &   29.2 &   136$\pm$3  &   2.5 &  169$\pm$3  &  32.21 &  125$\pm$ 3 &  2.0  \\
6731  [S{\sc ii}]            &  147$\pm$4   &   21.3 &    99$\pm$3  &   2.7 &  123$\pm$2  &  23.61 &   91$\pm$ 2 &  2.1  \\
7065  He{\sc i}              &   41$\pm$3   &    6.2 &    26$\pm$2  &   6.9 &   34$\pm$2  &   7.25 &   25$\pm$ 2 &  6.6  \\
7136  [Ar{\sc iii}]          &  127$\pm$4   &   20.5 &    80$\pm$3  &   3.3 &  102$\pm$3  &  22.14 &   72$\pm$ 2 &  2.8  \\
7254  O{\sc i}               &    2$\pm$1   &    0.4 &     1$\pm$1  &   9.2 &     ---     &   ---  &    ---      & ---   \\
7281  He{\sc i}$^a$          &    9$\pm$1   &    1.5 &     6$\pm$1  &   7.3 &    7$\pm$1  &   1.62 &    5$\pm$ 0 &  6.6  \\
7319  [O{\sc ii}]$^d$        &   35$\pm$1   &    5.5 &    21$\pm$1  &   3.1 &   29$\pm$1  &   6.30 &   20$\pm$ 1 &  2.5  \\
7330  [O{\sc ii}]$^e$        &   28$\pm$1   &    4.5 &    18$\pm$1  &   3.3 &   24$\pm$1  &   5.16 &   16$\pm$ 0 &  2.7  \\
7751  [Ar{\sc iii}]          &   34$\pm$2   &    5.8 &    20$\pm$1  &   6.1 &   25$\pm$1  &   5.81 &   16$\pm$ 1 &  5.0  \\
8392  P20                    &    9$\pm$1   &    2.0 &     5$\pm$1  &   7.6 &    7$\pm$1  &   2.16 &    4$\pm$ 0 &  8.1  \\
8413  P19                    &    9$\pm$1   &    2.1 &     5$\pm$1  &  11.0 &     ---     &   ---  &    ---      & ---   \\
8438  P18                    &    8$\pm$1   &    1.7 &     4$\pm$1  &  10.6 &    5$\pm$1  &   1.67 &    3$\pm$ 0 &  8.9  \\
8446  O{\sc i}               &   11$\pm$1   &    2.3 &     6$\pm$1  &   7.7 &   10$\pm$1  &   3.06 &    6$\pm$ 0 &  4.9  \\
8468  P17                    &    8$\pm$1   &    1.9 &     5$\pm$1  &   9.7 &    6$\pm$1  &   1.96 &    4$\pm$ 0 &  9.1  \\
8503  P16                    &   20$\pm$2   &    5.2 &    11$\pm$1  &  10.8 &     ---     &   ---  &    ---      & ---   \\
8546  P15                    &   13$\pm$1   &    3.5 &     7$\pm$1  &  11.1 &    9$\pm$1  &   3.46 &    6$\pm$ 1 &  9.2  \\
8599  P14                    &   12$\pm$1   &    2.7 &     7$\pm$1  &   9.3 &   13$\pm$1  &   4.07 &    8$\pm$ 1 &  9.1  \\

\hline

\end{tabular}
\end{center}
}
\end{table*}

\begin{table*}
{\small
\contcaption{Relative observed and reddening corrected emission line intensities [$F(H\beta)$=$I(H\beta)$=1000] for SDSS J162410.11-002202.5.}
\begin{center}
\begin{tabular}{|l|cccc|cccc|}
\hline
                           & \multicolumn{8}{c|}{SDSS J002101.03+005248.1 - spSpec-51900-0390-445} \\
\cline{2-9}
\multicolumn{1}{|c|}{$\lambda$  ({\AA})}  & \multicolumn{4}{c|}{WHT} & \multicolumn{4}{c|}{SDSS}   \\
 & $F(\lambda)$  &  -EW(\AA)  & $I(\lambda)$ & Error (\%) & $F(\lambda)$  & -EW(\AA)  & $I(\lambda)$ & Error (\%) \\
\hline

8665  P13                    &   15$\pm$1   &    3.6 &     8$\pm$1  &   8.3 &   16$\pm$1  &   5.73 &   10$\pm$ 1 &  7.5  \\
8751  P12                    &   23$\pm$2   &    6.0 &    12$\pm$1  &   9.2 &   18$\pm$2  &   5.94 &   11$\pm$ 1 &  8.7  \\
8865  P11                    &   33$\pm$3   &    9.0 &    17$\pm$2  &  10.4 &   24$\pm$2  &   8.24 &   15$\pm$ 1 & 10.1  \\
9014  P10                    &   34$\pm$6   &    8.3 &    17$\pm$3  &  17.6 &     ---     &   ---  &    ---      & ---   \\
9069  [S{\sc iii}]$^f$       &  137$\pm$9   &   66.8 &    70$\pm$5  &   7.0 &     ---     &   ---  &    ---      & ---   \\
9532  [S{\sc iii}]           &  809$\pm$36  &  220.0 &   401$\pm$19 &   4.8 &     ---     &   ---  &    ---      & ---   \\
9547  P8                     &   91$\pm$3   &   25.8 &    45$\pm$2  &   4.1 &     ---     &   ---  &    ---      & ---   \\

\hline
I(H$\beta$)(erg\,seg$^{-1}$\,cm$^{-2}$) & \multicolumn{4}{c|}{5.03\,$\times$\,10$^{-14}$} &  \multicolumn{4}{c|}{8.10\,$\times$\,10$^{-14}$} \\
C(H$\beta$) & \multicolumn{4}{c|}{0.52$\pm$0.01} &  \multicolumn{4}{c|}{0.39$\pm$0.01} \\
\hline

\end{tabular}
\end{center}

\medskip
\begin{flushleft} 
$^a$\,possibly blend with an unknown line; $^b$\,[O{\sc ii}]
$\lambda\lambda$\,3726\,+\,3729;  $^c$\,[Fe{\sc iii}]
$\lambda\lambda$\,4986\,+\,4987; $^d$\,[O{\sc ii}]
$\lambda\lambda$\,7318\,+\,7320; $^e$\,[O{\sc ii}]
$\lambda\lambda$\,7330\,+\,7331;  $^f$\,affected by atmospheric absorption bands.
\end{flushleft} 
}
\end{table*}

\section{Physical conditions of the gas}
\label{physi}

\subsection{Electron densities and temperatures from forbidden lines}

\nocite{1987JRASC..81..195D}
\nocite{1995PASP..107..896S}
\nocite{1982MNRAS.198..111Z}
\nocite{1976MNRAS.177...31P}
\nocite{2004A&A...427..873W}
\nocite{1992AJ....103.1330G}
\nocite{1999ApJ...526..544T}
\nocite{2004ApJS..153..429K}
\nocite{1980ApJ...240...99B}
\nocite{1994ApJ...435..647I}
\nocite{2004A&A...415...87I}
\nocite{1990A&AS...83..501S}

The physical conditions of the ionized gas, including electron temperatures and
electron density, have been derived from the emission line data using the same
procedures as in P\'erez-Montero \& D\'\i az (2003), based on the five-level
statistical equilibrium atom approximation in the task TEMDEN, of the software
package IRAF 
(De Robertis, Dufour \& Hunt, 1987; Shaw \& Dufour, 1995). The atomic
coefficients used here are the  same as in P\'erez-Montero \& D\'\i az (2003;
see table 4 of that work), except in the case of O$^+$ for which we have used
the transition probabilities from Zeippen (1982) and the collision strengths
from Pradham (1976), which offer more reliable nebular diagnostics results for
this species (Wang et al., 2004). We have taken as sources of error the
uncertainties associated with the measurement of the emission-line fluxes and
the reddening correction and we have propagated them through our calculations. 
Electron densities have been derived from the [S{\sc
    ii}]\,$\lambda\lambda$\,6717\,/\,6731\,\AA\ line ratio, which is
representative of the low-excitation zone of the ionized gas. { In all 
cases they were found to be lower than 200 particles per cubic centimeter,
well below the critical density for collisional deexcitation.}
We have tried to derive the electron densities from the [Ar{\sc
    iv}]\,$\lambda\lambda$\,4713\,/\,4740\,\AA\ line ratio, by decontaminating
the first one from the He{\sc i} contribution, but the derived density values
had unacceptable errors due to their large and sensitive dependencies on the
emission line intensities and the errors of the observed fluxes. 

Several electron temperatures for each spectrum have been measured:
T([O{\sc ii}]), T([O{\sc iii}]), T([S{\sc ii}]), T([S{\sc iii}]) and T([N{\sc
    ii}]). The [N{\sc ii}] line at 5755 \AA\ was not detected in the spectrum of SDSS
J003218.60+150014.2 and therefore it was not possible to measure T([N{\sc ii}])
for this object. Both the [O{\sc ii}]\,$\lambda\lambda$\,7319,7330\,\AA\ and the
[N{\sc ii}]\,$\lambda$\,5755\,\AA\  lines can have a contribution by direct
recombination which increases with temperature. Using the calculated [O{\sc
    iii}] electron temperatures, we have estimated these contributions to be
less than 4\% in all cases and therefore we have not corrected for this
effect, although we have considered them when  estimating the errors. The
emission-line ratios used to calculate each temperature are summarized in table
\ref{ratios}.  

\begin{table}
\begin{minipage}{85mm}
\vspace{-0.3cm}
\normalsize
\caption{Emission-line ratios used to derive electron densities and temperatures.}
\begin{center}
\begin{tabular}{cc}
\hline
\hline
Diagnostic  & Lines  \\
\hline
n([S{\sc ii}])     &     I(6717\AA)/I(6731\AA) \\
T([O{\sc iii}])    &    (I(4959\AA)+I(5007\AA))/I(4363\AA) \\
T([O{\sc ii}])     &     I(3727\AA)/(I(7319\AA)+I(7330\AA)) \\
T([S{\sc iii}])    &    (I(9069\AA)+I(9532\AA))/I(6312\AA) \\
T([S{\sc ii}])     &    (I(6717\AA)+I(6731\AA))/(I(4068\AA)+I(4074\AA)) \\
T([N{\sc ii}])     &    (I(6548\AA)+I(6584\AA))/I(5755\AA) \\
\hline
\end{tabular}
\end{center}
\label{ratios}
\end{minipage}
\end{table}

\begin{table*}
{\small
\caption{Electron densities and temperatures for the observed
galaxies using SDSS and WHT spectroscopy. Temperatures marked with asterisks
have been deduced using equations from grids of photoionization models.} 
\label{temden}
\begin{center}
\begin{tabular}{l cc cc cc }
\hline
\hline
               & \multicolumn{2}{c}{SDSS J002101.03+005248.1} & \multicolumn{2}{c}{SDSS J003218.60+150014.2} &   \multicolumn{2}{c}{SDSS J162410.11-002202.5}\\
               &   WHT          &      SDSS        &   WHT            &  SDSS            &   WHT         & SDSS              \\
\hline                                    
n([S{\sc ii}]) &   120$\pm$68   &   77$\pm$58      &   54:            &   56:            &   58:         &   66:       \\
T([O{\sc iii}])&  1.25$\pm$0.02 & 1.13$\pm$0.02    & 1.28$\pm$0.02    & 1.28$\pm$0.03    & 1.24$\pm$0.01 & 1.16$\pm$0.01     \\ 
T([O{\sc ii}]) &  1.03$\pm$0.02 & 1.05$\pm$0.02    & 1.35$\pm$0.04    &   ---            & 1.31$\pm$0.03 & 1.23$\pm$0.03     \\ 
$\langle$T([O{\sc ii}])$\rangle^*$ &  --- & --- & --- & 1.36 & --- & ---   \\ 
T([S{\sc iii}])&  1.31$\pm$0.05 & ---              & 1.36$\pm$0.07    &   ---            & 1.26$\pm$0.04 &     ---           \\ 
$\langle$T([S{\sc iii}])$\rangle^*$ & --- & 1.02 & ---   & 1.21    & --- & 1.06    \\  
T([S{\sc ii}]) &  0.86$\pm$0.06 & 0.76$\pm$0.06    & 1.03$\pm$0.05    & 0.92$\pm$0.06    & 1.04$\pm$0.07 & 1.02$\pm$0.09     \\ 
T([N{\sc ii}]) &  1.19$\pm$0.05 & 1.36$\pm$0.06    &   ---            &   ---            & 1.42$\pm$0.08 &     ---           \\ 
$\langle$T([N{\sc ii}])$\rangle^*$ & --- & --- & 1.35   & 1.36   & --- & 1.23    \\ 
T(Bac) & 1.24$\pm$0.27 & --- & 0.96$\pm$0.16    & ---    & 1.23$\pm$0.22 & ---    \\ 
T(H$\beta$)   & 1.24$\pm$0.31 & --- & 1.02$\pm$0.20    & ---    & 1.24$\pm$0.26 & ---    \\ 
T(He{\sc ii}) & 1.24$\pm$0.29 & --- & 0.98$\pm$0.19    & ---    & 1.24$\pm$0.23 & ---    \\ 
\hline
\multicolumn{7}{l}{densities in $cm^{-3}$ and temperatures in 10$^4$\,$\degr$K}
\end{tabular}
\end{center}}
\end{table*}


The derived electron densities and temperatures for the three observed objects
are listed in columns 2, 4 and 6 of table \ref{temden} along with their
corresponding errors.

\subsection{Balmer temperature}
\label{balmer}

\begin{figure}
\includegraphics[width=.37\textwidth,angle=270]{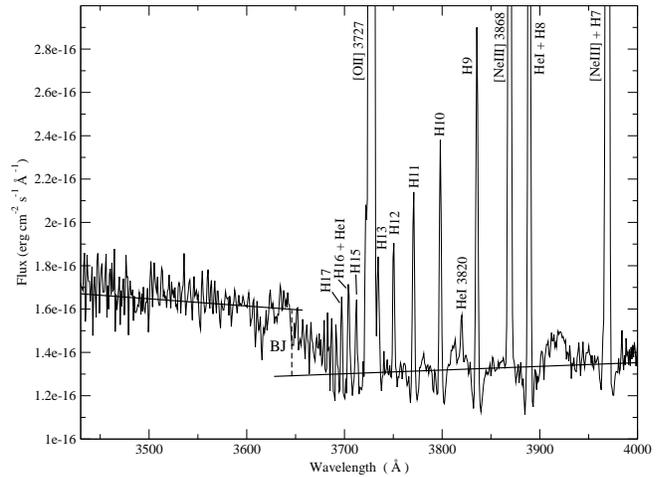}
\caption{Enlargement of the spectrum of SDSS J003218.60+150014.2 taken with the
  WHT. Its spectral range is from 3430 to 4000\,\AA, and it is in the rest frame.
  The solid lines trace the continuum to both sides of the Balmer jump and the
  dashed line depict the value of its measurement.} 
\label{Bjump}
\end{figure}

The Balmer temperature depends on the value of the Balmer jump (BJ) in
emission. To measure this value we have adjusted the continuum at both sides of
the discontinuity ($\lambda_B$\,=\,3646\,\AA). Fig. \ref{Bjump} shows an example
of the procedure for  SDSS J003218.59+150014.2. The contribution of the
underlying population (see Section 3) affect, among  others, the hydrogen
emission lines near the Balmer jump. The increase of the number of lines toward
shorter wavelengths produces blends which tend to depress the continuum level to
the right of the discontinuity { and precludes the application of
multi-gaussian component fittings}. We have taken special care in the definition
of this continuum by using a long baseline { and spectral windows free of
absorption lines}.  The uncertainties due to the 
presence of the underlying stellar population (different possible continuum
placements) have been included in the errors of the measurements of the
discontinuities. { They are actually smaller than the error introduced by the
fitting of stellar templates.} Once the Balmer jump is measured, the Balmer
continuum 
temperature (T(Bac)) is determined from the ratio of the Balmer jump flux to the
flux of the H11 Balmer emission line using equation (3) in
\cite{2001MNRAS.327..141L}: 
\[
T(Bac)\,=\,368\times(1\,+\,0.259y^+\,+\,3.409y^{2+})\Big(\frac{BJ}{H11}\Big)^{-3/2}\,K
\]
\noindent
where $y^+$ and $y^{2+}$ are the ionic abundances of singly and doubly ionized
helium, He$^+$/H$^+$ and He$^{2+}$/H$^+$ (see \S \ref{abund}), respectively,
and BJ is in ergs\,cm$^{-2}$\,s$^{-1}$\,\AA$^{-1}$. The derived values for
T(Bac) are also listed in table \ref{temden}.

\section{Chemical abundances}
\label{abund}
Ionic and total abundances of He, O, S, N, Ne, Ar and Fe are obtained as detailed
below and given in tables \ref{absHe} and \ref{abs}.
We have derived the ionic chemical abundances of the different species using the stronger
available emission lines detected in the analyzed spectra. The total abundances
have been derived by taking into account, when required, the unseen ionization stages of each
element, resorting to the most widely accepted ionization correction factors
(ICF) for each species: 
\[\frac{X}{H} = ICF(X^{+i}) \cdot \frac{X^{+i}}{H^+}\]

\subsection{Helium abundance}

\begin{figure}
\includegraphics[width=.36\textwidth,angle=270]{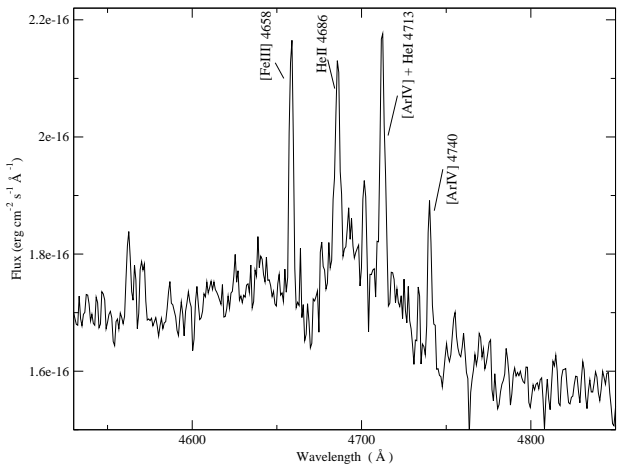}\\
\caption{Wolf-Rayet features in SDSS J162410.11-002202.5.}
\label{WRbumps}
\end{figure}

We have measured emission fluxes for 10 lines of He{\sc i} and 1 of
He{\sc ii}, although four of them are blended with another emission
line and two are so weak that they cannot be used to derive the helium
abundance with the necessary accuracy. We have therefore used the He{\sc
  i}\,$\lambda\lambda$\,4471, 5876, 6678 and 7065\,\AA, and He{\sc
  ii}\,$\lambda$\,4686\,\AA\ lines to estimate the abundances of helium once and
twice ionized respectively. These lines arise mainly from pure recombination,
however they could have some contribution from collisional excitation as well as
be affected by self-absorption and, if present, by underlying stellar
absorption \citep[see][for a complete treatment of these
  effects]{2001NewA....6..119O, 2004ApJ...617...29O}. We have taken the electron
temperature  
of [O{\sc iii}] as representative of the zone where the He emission arises
(i. e. T(He{\sc ii})\,$\simeq$\,T([O{\sc iii}])) and we have used the equations given
by Olive \& Skillman  to derive the He$^{+}$/H$^{+}$ 
value, using the theoretical emissivities scaled to H$\beta$ from
\cite{1996MNRAS.278..683S} and the expressions for the collisional correction
factors from \cite{1995ApJ...442..714K}. We have not taken into account, however,
the corrections for fluorescence (three of the used helium lines have a negligible
dependence with optical depth effects and the observed  objects have low densities) and
the underlying stellar population.
The three galaxies show in their spectra the signature of the presence of  WR
stars by the blue `bump' around $\lambda$\,4600\,\AA\,\ therefore we have to take
care when measuring the emission line flux of He{\sc ii}\,$\lambda$\,4686\,\AA\
(see Fig. \ref{WRbumps}).
We have used equation (9) from \cite{1983ApJ...273...81K} to
calculate the abundance of twice ionized helium.
Then, for the total abundance we have directly added the two  ionic
abundances. 
\[
\frac{He}{H}\,=\,\frac{He^++He^{2+}}{H^+}
\]
\noindent The results obtained for each line and their corresponding errors are
presented in table \ref{absHe}, along with the adopted value for
He$^{+}$/H$^{+}$.

\begin{table*}
\begin{minipage}{170mm}
{\small
\vspace{-0.3cm}
\caption{Ionic and total chemical abundances for helium.} 
\label{absHe}

\begin{center}
\begin{tabular}{cc| cccccc}
\hline
\hline
     & & \multicolumn{2}{c}{SDSS J002101.03+005248.1} & \multicolumn{2}{c}{SDSS J003218.60+150014.2}&   \multicolumn{2}{c}{SDSS J162410.11-002202.5}  \\
                  & $\lambda$\,(\AA)   &      WHT      &        SDSS   &     WHT       &        SDSS   &   WHT         &      SDSS       \\
\hline

$He^+/H^+$        & 4471 & 0.086$\pm$0.006 & 0.077$\pm$0.006 & 0.076$\pm$0.004 & 0.074$\pm$0.006 & 0.086$\pm$0.003 & 0.079$\pm$0.002   \\
                  & 5876 & 0.096$\pm$0.005 & 0.094$\pm$0.006 & 0.096$\pm$0.004 & 0.088$\pm$0.003 & 0.104$\pm$0.002 & 0.086$\pm$0.002   \\
                  & 6678 & 0.087$\pm$0.004 & 0.089$\pm$0.006 & 0.089$\pm$0.004 & 0.079$\pm$0.001 & 0.094$\pm$0.005 & 0.091$\pm$0.005   \\
                  & 7065 & 0.094$\pm$0.005 & 0.102$\pm$0.006 & 0.086$\pm$0.009 & 0.076$\pm$0.006 & 0.104$\pm$0.008 & 0.099$\pm$0.007   \\
               & adopted & 0.090$\pm$0.005 & 0.090$\pm$0.009 & 0.087$\pm$0.007 & 0.080$\pm$0.005 & 0.098$\pm$0.007 & 0.085$\pm$0.009   \\
       
$He^{2+}/H^+$     & 4686 & 0.0006$\pm$0.0001 & 0.0004$\pm$0.0001 & 0.0011$\pm$0.0001 & 0.0013$\pm$0.0001 & 0.0008$\pm$0.0001 & 0.0007$\pm$0.0001   \\
       
\bf{(He/H)}       &    & 0.091$\pm$0.005 &  0.090$\pm$0.009 & 0.088$\pm$0.007 & 0.081$\pm$0.005 & 0.098$\pm$0.007 & 0.085$\pm$0.009   \\
\hline
\end{tabular}
\end{center}}
\end{minipage}
\end{table*}


\begin{table*}
\begin{minipage}{170mm}
{\small
\vspace{-0.3cm}
\caption{Ionic and total chemical abundances, along with their corresponding
  ionization correction factors, derived from CELs. Logarithm of the S$_{23}$
  parameter.}
\label{abs}

\begin{center}
\begin{tabular}{c|cccccc}
\hline
\hline
     & \multicolumn{2}{c}{SDSS J002101.03+005248.1} & \multicolumn{2}{c}{SDSS J003218.60+150014.2}&   \multicolumn{2}{c}{SDSS J162410.11-002202.5}  \\
                      &      WHT      &        SDSS   &     WHT       &        SDSS   &   WHT         &      SDSS       \\
\hline

12+$\log(O^+/H^+)$    & 7.73$\pm$0.06 & 7.73$\pm$0.05 & 7.15$\pm$0.06 & 7.10 & 7.18$\pm$0.06 &  7.30$\pm$0.05  \\
12+$\log(O^{2+}/H^+)$ & 7.86$\pm$0.02 & 8.01$\pm$0.02 & 7.86$\pm$0.02 & 7.87$\pm$0.05 & 7.98$\pm$0.02 &  8.08$\pm$0.01  \\
\bf{ 12+log(O/H)}     & 8.10$\pm$0.04 & 8.19$\pm$0.03 & 7.93$\pm$0.03 & 7.93 & 8.05$\pm$0.02 &  8.14$\pm$0.02  \\
\hline                                                                  
12+$\log(S^+/H^+)$    & 5.93$\pm$0.11 & 6.13$\pm$0.12 & 5.80$\pm$0.06 & 5.88$\pm$0.08 & 5.69$\pm$0.08 &  5.67$\pm$0.10  \\
12+$\log(S^{2+}/H^+)$ & 5.91$\pm$0.05 & 6.55 & 6.16$\pm$0.06 & 6.34 & 6.14$\pm$0.04 &  6.49  \\
ICF($S^++S^{2+}$)     & 1.12$\pm$0.10 & 1.18 & 1.50$\pm$0.15 & 1.57 & 1.61$\pm$0.13 &  1.58  \\
\bf{ 12+log(S/H)}     & 6.27$\pm$0.12 & 6.77 & 6.49$\pm$0.11 & 6.67 & 6.48$\pm$0.09 &  6.75  \\
\hline                                                                  
12+$\log(N^+/H^+)$    & 6.68$\pm$0.04 & 6.62$\pm$0.03 & 6.03 & 5.94 & 5.92$\pm$0.06 &  5.98  \\
\bf{log(N/O)}         & -1.06 $\pm$0.10 & -1.12 $\pm$0.08 & -1.11 & -1.16 & -1.26$\pm$0.12 &  -1.32  \\
\hline                                                                  
12+$\log(Ne^{2+}/H^+)$& 7.27$\pm$0.03 & 7.42$\pm$0.03 & 7.30$\pm$0.06 & 7.27$\pm$0.05 & 7.33$\pm$0.03 &  7.44$\pm$0.02  \\
\bf{log(Ne/O)}        & -0.59$\pm$0.06 & -0.59$\pm$0.06 & -0.63$\pm$0.06 & -0.60$\pm$0.07 & -0.65$\pm$0.05 &  -0.64$\pm$0.03  \\
\hline                                                                  
12+$\log(Ar^{2+}/H^+)$& 5.50$\pm$0.05 & 5.67 & 5.65$\pm$0.06 & 5.62 & 5.66$\pm$0.04 &  5.70  \\
12+$\log(Ar^{3+}/H^+)$& 4.37$\pm$0.05 &   ---         & 4.32$\pm$0.07 & 4.37$\pm$0.05 & 4.36$\pm$0.05 &  4.43$\pm$0.05  \\
ICF($Ar^{2+}$+$Ar^{3+}$)& 1.21$\pm$0.06 & ---         & 1.01$\pm$0.01 & 1.02 & 1.01$\pm$0.01 &  1.02$\pm$0.01  \\
\bf{ 12+log(Ar/H)}    & 5.59$\pm$0.08 & --- & 5.66$\pm$0.06 & 5.63 & 5.66$\pm$0.05 &  5.70  \\
\hline                                                                  
12+$\log(Fe^{2+}/H^+)$& 5.50$\pm$0.05 & 5.66$\pm$0.07 & 5.42$\pm$0.06 & 5.46$\pm$0.06 & 5.48$\pm$0.05 &  5.59$\pm$0.05  \\
ICF($Fe^{2+}$)        & 3.09$\pm$0.33 & 3.52$\pm$0.32 & 6.04$\pm$0.84 & 6.63 & 7.00$\pm$0.83&  6.72$\pm$0.63  \\
\bf{ 12+log(Fe/H)}    & 5.99$\pm$0.10 & 6.20$\pm$0.11 & 6.20$\pm$0.12 & 6.28 & 6.32$\pm$0.10 &  6.42$\pm$0.09  \\
\hline
\bf{log(S$_{23}$)}    & -0.22$\pm$0.02 & ---  & -0.02$\pm$0.03  & ---   & -0.10$\pm$0.02  & ---  \\
\hline
\end{tabular}
\end{center}}
\end{minipage}
\end{table*}


\subsection{Ionic and total chemical abundances from forbidden lines}

The oxygen ionic abundance ratios, O$^{+}$/H$^{+}$ and O$^{2+}$/H$^{+}$, have
been derived from the [O{\sc ii}]\,$\lambda\lambda$\,3727,29\,\AA\ and [O{\sc
    iii}] $\lambda\lambda$\,4959, 5007\,\AA\ lines respectively. The
simultaneous determination of T([O{\sc ii}]) and T([O{\sc iii}]) allows a more
confident estimate of the total abundance of oxygen for which we have used the
approximation:  
\[ 
\frac{O}{H}\,=\,\frac{O^++O^{2+}}{H^+}
\]

Regarding sulphur, we have derived S$^+$ abundances using T([S{\sc ii}]) values and
the fluxes of the [S{\sc ii}] emission lines at $\lambda\lambda$\,6717, 6731\,{\AA}. 
In the same way, the abundances of
S$^{2+}$ have been derived from the values of the directly
measured T([S{\sc iii}]) and the near-IR [S{\sc iii}]\,$\lambda\lambda$\,9069, 9532\,\AA\
lines. The total sulphur abundance has been calculated using an ICF for S$^+$+S$^{2+}$
according to Barker's (1980) formula, { which is based on the photoionization
models by \cite{1978A&A....66..257S}}: 
 
\[
ICF(S^++S^{2+}) = \left[ 1-\left( 1-\frac{N(O^+)}{N(O)}
  \right)^\alpha\right]^{-1/\alpha} 
\]
\noindent where $\alpha$\,=\,2.5 gives the best fit to the scarce observational
data on S$^{3+}$ abundances (P\'erez-Montero et al.\ 2006). 

The ionic abundance of nitrogen, N$^{+}$/H$^{+}$ has been derived from the
intensities of the $\lambda\lambda$\,6548, 6584\,\AA\ lines and  the measured
electron temperature of [N{\sc ii}]  for objects SDSS J002101.03+005248.1 and
SDSS J162410.11-002202.5. Then, the N/O abundance has been derived under the
assumption that  
\[
\frac{N}{O}\,=\,\frac{N^+}{O^+}
\]
In the case of SDSS J003218.60+150014.2, the [N{\sc
    ii}]\,$\lambda$\,5755\,\AA\ line could not be measured and therefore
we have assumed that T([N{\sc ii}]) is equal to T([O{\sc ii}]) and we have derived
directly the N/O ratio according to the expression given in
    \cite{1992MNRAS.255..325P}.

Neon is only visible in all the spectra via the [Ne{\sc iii}] emission line at
$\lambda$3868\,{\AA}. For this ion we have taken the electron temperature 
of [O{\sc iii}], as
representative of the high excitation zone. The total abundance of neon has been
calculated assuming that:
\[
\frac{Ne}{O}\,=\,\frac{Ne^{2+}}{O^{2+}}
\]
Izotov et al.\ (2004) point out that this assumption can lead to an overestimate of
Ne/H in objects with low excitation, where the charge transfer between O$^{2+}$
and H$^0$ becomes important. Nevertheless, in our case, it is probably justified
given the high excitation of the observed  objects.
 
The main ionization states of Ar in ionized regions are  Ar$^{2+}$ and
Ar$^{3+}$. The abundance of Ar$^{2+}$ has been calculated by means of the
emission line of 
[Ar{\sc iii}]\,$\lambda$\,7136\,\AA\ assuming that T([Ar{\sc iii}])\,$\approx$\,T([S{\sc iii}]) 
(Garnett, 1992), while the  ionic abundance of Ar$^{3+}$ has been calculated under
the assumption that T([Ar{\sc iv}])\,$\approx$\,T([O{\sc iii}]) and 
using the emission line of [Ar{\sc iv}]\,$\lambda$\,4740\,\AA.
We could have used the blended emission line [Ar{\sc iv}]+He{\sc
  I}\, at $\lambda$\,4713\,\AA\  subtracting the helium contribution. However, due
to the relative abundance of these species and the signal-to-noise ratio of this
blended line, we prefer not to estimate this ionic abundance with such a  large
error. The total abundance of Ar has been calculated  using the
ICF(Ar$^{2+}$+Ar$^{3+}$) given by Izotov et al.\ (1994) which, in turn, has been
derived from the photo-ionization models by Stasi\'nska (1990) as: 
\begin{eqnarray*}
\label{tcel}
\lefteqn{ICF(Ar^{2+}+Ar^{3+})\,=\,\Big[0.99+0.091\Big(\frac{O^+}{O}\Big)
{} } \nonumber\\ & & {} 
-1.14\Big(\frac{O^+}{O}\Big)^2+0.077\Big(\frac{O^+}{O}\Big)^3\Big]^{-1}
{}
\end{eqnarray*}

Finally, for iron we have used the emission line of [Fe{\sc
    iii}]\,$\lambda$\,4658\,\AA\ and the electron temperature of
[O{\sc iii}]. We have taken the ICF(Fe$^{2+}$) from \cite{2004IAUS..217..188R},
which yields:  
\[
ICF(Fe^{2+})\,=\,\Big(\frac{O^+}{O^{2+}}\Big)^{0.09} \cdot
\Big[1+\frac{O^{2+}}{O^{+}}\Big]
\]

The ionic and total abundances for each observed element are presented in
columns 2, 4 and 6 of table \ref{abs}, along with their corresponding errors.

\section{Discussion}

\subsection{Comparison with SDSS data}

In order to compare our results with those provided by SDSS spectra, we have
measured the emission line intensities and equivalent widths on the SLOAN
spectra of
the three observed objects in the same way as described in Section 3. In order
to allow an easy comparison between the results on both sets of spectra, we
have listed these values in columns 6 to 9 of tables \ref{inten1}, \ref{inten2},
and \ref{inten3}.  

Strong emission line fluxes relative to H$\beta$ measured on WHT and SDSS
spectra differ by less than 10\% for SDSS J003218.60+150014.2 and SDSS
J162410.11-002202.5 and about 30\% for SDSS J002101.03+005248.1. This is
partially compensated by differences in the derived reddening constant so that
reddening corrected emission line intensities relative to H$\beta$ differ by
less that 10\% for the [O{\sc ii}]\,$\lambda\lambda$\,3727,29\,\AA\ line, about
15\% for the weak  [O{\sc iii}] $\lambda$\,4363\,\AA\ and  [S{\sc
    iii}]\,$\lambda$\,6312\,\AA\ lines and only a few percent for the
strong  [O{\sc iii}]\,$\lambda$\,5007\,\AA\ line.  In fact, given the
difference in aperture between both sets of 
observations, 0.5 and 3 arcsec for WHT and SDSS respectively, some differences
are to be expected. While we have probably observed the bright cores of the
galaxies where most of the light and present star formation is concentrated, the 
SDSS observations map a more extensive area which could include external
diffuse zones. This is evidenced by the more conspicuous underlying stellar
population detected in the WHT spectra which leads to lower values of the
emission line equivalent widths. 
The best agreement between the two sets of measurements is found for SDSS
J003218.60+150014.2, which is probably the more compact object, as evidenced by
the difference in the measured H$\beta$ fluxes which is only a factor of 2 and
the close agreement between the measured equivalent widths on the two spectra. 

\begin{figure}
\includegraphics[width=8.5cm,angle=270,clip=]{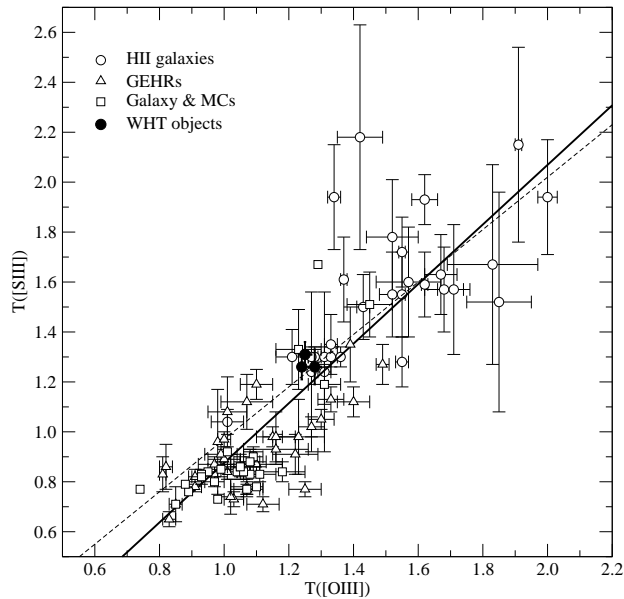}\\
\caption{{ Relation between T([S{\sc iii}]) and T([O{\sc iii}]) for the observed
  objects (solid circles), and \HII\ galaxies (open circles), Giant Extragalactic
  \HII\ regions (upward triangles) and diffuse \HII\ regions in the Galaxy and
  the Magellanic Clouds (squares) , for which data on the auroral and nebular lines of
  [O{\sc iii}] and [S{\sc iii}]  exist (see P\'erez-Montero et al.\ 2006). The
  temperatures are in units of 10$^4$\,$\degr$K.}} 
\label{T([SIII]-T([OIII])}
\end{figure}

The SDSS spectra have been analyzed following the same methodology as explained
in Sections 4 and 5 for the derivation of temperatures and abundances although,
due to the different nature of the observations: different spectral coverage,
signal-to-noise ratio and spectral resolution, some further assumptions had to
be made. These refer mainly to the temperatures of the different ions.  
Regarding sulphur, the SDSS data do not reach the 9000-9600\,\AA\ range covered
by the WHT spectra and therefore it was not possible to determine directly
T([S{\sc iii}]). The relation between T([S{\sc iii}]) and T([O{\sc iii}]) is
reproduced in Fig. \ref{T([SIII]-T([OIII])}. 
{ The sample used for comparison is a compilation of published data for which
measurements of the nebular and auroral lines of [O{\sc iii}] and [S{\sc iii}]
exist, thus allowing the simultaneous determination of T([O{\sc iii}]) and
T([S{\sc iii}]). The sample is listed in Tables 1 and 2 of P\'erez-Montero et
al. (2006; objects not marked with a superscript b, as explained in the text),
where the reader can find all the information concerning the data.} 
The dashed line in the plot corresponds to the theoretical
relation based on the grids of photoionization models described in
P\'erez-Montero \& D\'\i az (2005),  
\[
T([S\textrm{\sc iii}])\,=\,1.05\,T([O\textrm{\sc iii}])\,-\,0.08
\]
\noindent which  differs slightly from the semi-empirical relation by Garnett
(1992) mostly due to the introduction of the new atomic coefficients for
S$^{2+}$ from Tayal \& Gupta (1999). 
The solid line in Fig. \ref{T([SIII]-T([OIII])} corresponds to the actual fit to
the data: 
\[T([S\textrm{\sc iii}]) = (1.19 \pm 0.08)\,T([O\textrm{\sc iii}]) - (0.32 \pm 0.10)\]
The individual errors have not been taken into account in performing the fit. 
{ This fit is different from that found by Garnett (1992) that seems to
reproduce well the M101 \HII\ region data analyzed by Kennicutt, Bresolin \&
Garnett (2003; KBG03). This is mostly due to the larger temperature baseline
that we use. The object with the highest T([S{\sc iii}]) in KBG03 is NGC5471A
(12800\,$\degr$K); our sample includes high excitation \HII\ galaxies with
T([S{\sc iii}]) up to 24000\,$\degr$K, while including at the same time KBG03
sample. The introduction of 
the high excitation objects make the relation steeper and increases the error
of the calibration. This illustrates the danger of extrapolating relations
found for a restricted range of values. }
We have used our empirical calibration in order to obtain T([S{\sc
iii}]) for the SDSS spectra. The estimated errors  introduced by the
calibration are of the order of 12\% for T([S{\sc iii}]),  i.e. between 1400 and
1500\,$\degr$K for the observed objects.  

\nocite{2003ApJ...591..801K}


\begin{figure}
\includegraphics[width=8.5cm,angle=270,clip=]{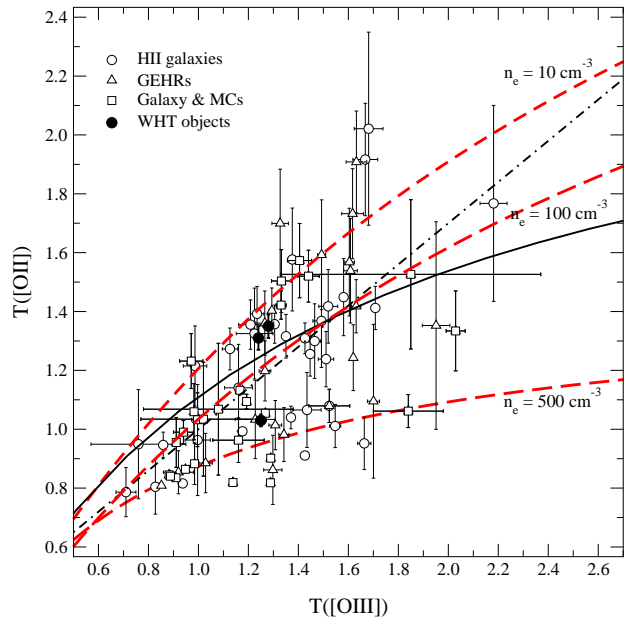}\\
\caption{{ Relation between T([O{\sc ii}]) and T([O{\sc iii}]) for the observed
  objects (solid circles) and \HII\ galaxies (open circles), Giant Extragalactic
  \HII\ regions (upward triangles) and diffuse \HII\ regions in the Galaxy and
  the Magellanic Clouds (squares) from P\'erez-Montero \& D\'\i az (2005). The 
  dashed lines correspond to photoionization models from P\'erez-Montero \&
  D\'\i az (2003) for electron densities n$ _{e} $ = 10, 100 and 500
  cm$^{-3}$. The model sequences from Stasi\'nska (1980; solid line) and
  Stasi\'nska (1990; dashed-dotted line) are also shown. The temperatures are in units of
  10$^4$\,$\degr$K.}}
\label{T([OII]-T([OIII])}
\end{figure}

Likewise, the SDSS spectrum of J003218.60+150014.2 does not include the lines of
[O{\sc ii}] at $\lambda\lambda$\,3727,29\,\AA\ and therefore it was not possible
to derive T([O{\sc ii}]). Again, we have resorted to the model predicted
relationship between  T([O{\sc ii}]) and T([O{\sc iii}]) found by 
P\'erez-Montero \& D\'iaz (2003) that takes explicitly into account the
dependence of T([O{\sc ii}]) on electron density. { Three model sequences are
represented in Fig. \ref{T([OII]-T([OIII])} correspondeing to three different
values of the density: 10, 100 and 500 cm$ ^{-3} $ (dashed lines in the
plot). The model sequence for n= 100 cm$ ^{-3} $ is very similar to the one
derived from the models by Stasi\'nska (1980). The sample used for comparison
comprises the objects from P\'erez-Montero \& D\'\i az (2005) for which the
derivation of T([O{\sc ii}])
and T([O{\sc iii}]) has been possible.  In this case, due to the dependence of
T([O{\sc ii}]) on electron density, there is not a single empirical calibration
and it is not possible to give an estimate of 
the error introduced by the application of this procedure.}
 

\nocite{1980A&A....84..320S}

In the case of nitrogen, the [N{\sc ii}]\,$\lambda$\,5755\,\AA\ line could be
measured only in the SDSS spectrum of J002101.03+005248.1, due to poor signal to
noise. In the other two cases, the assumption T([N{\sc ii}]) = T([O{\sc ii}])
has been made. This assumption is usually made in standard analysis techniques;
however, there are not enough data for \HII\ galaxies to test it empirically. 

Finally, the Balmer continuum temperature could not be calculated from the SDSS
spectra due to lack of spectral coverage. 

Concerning  abundances, the S$^{2+}$/H$^{+}$ abundance ratios had to be
calculated using the intensity of the weak auroral [S{\sc iii}] line at
$\lambda$\,6312\,\AA\ and, in the case of SDSS J003218.60+150014.2, the
O$^{+}$/H$^{+}$ abundance ratio was derived using the [O{\sc
    ii}]\,$\lambda\lambda$\,7319,7330\, \AA\ lines following the procedure
described by Kniazev et al. (2004).  

The values of electron density and temperatures, ionic and total abundances
derived from the SDSS spectra are listed in columns 3, 5 and 7 of table
\ref{abs}, along with their corresponding errors in the cases where they have
been derived from measured emission lines intensities. Otherwise, since the
uncertainties introduced by the different assumptions made and the theoretical
models used are impossible to quantify, no formal errors are given. These
quantities should be considered as estimates and be used with caution. 

For SDSS J003218.60+150014.2 the values we have obtained for densities,
temperatures and abundances from the WHT and SDSS spectra 
are in excellent agreement within the observational errors, as expected from the
close agreement between the measured emission line intensities. For the other
two objects, the agreement can be considered as satisfactory, taking into
account the difference in aperture between both sets of observations.

\subsection{Gaseous physiscal conditions and element abundances}

\begin{figure}
\includegraphics[width=.48\textwidth,angle=0,clip=]{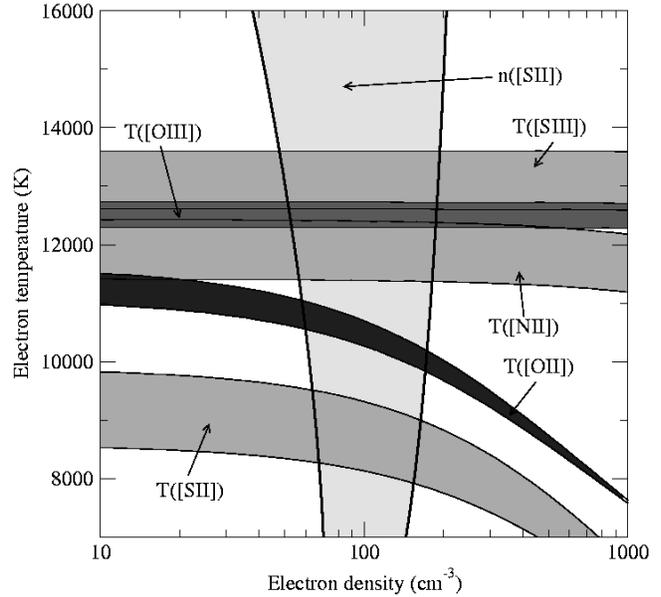}
\caption{Measured line temperatures and electron density for object SDSS
  J002101.03+005248.1. The width of the bands correspond to one $\sigma$ error.} 
\label{temperaturas0390}
\end{figure}

\begin{figure}
\includegraphics[width=.48\textwidth,angle=0,clip=]{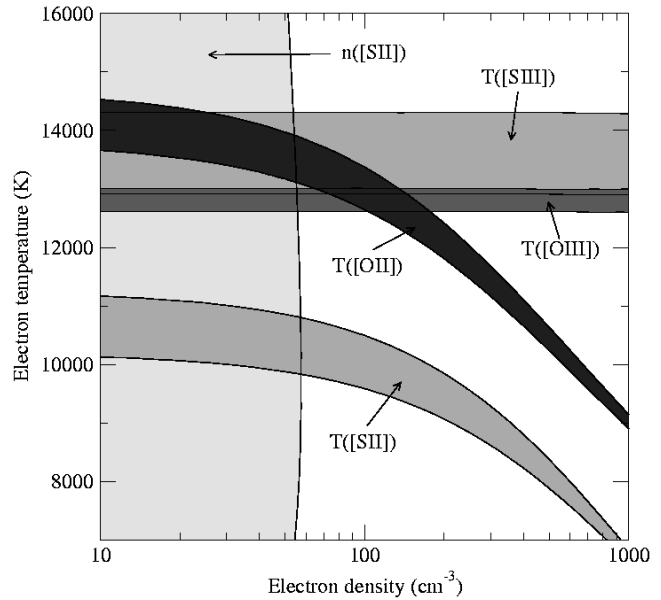}
\caption{Measured line temperatures and electron density for object SDSS
  J003218.60+150014.2. The width of the bands correspond to one $\sigma$ error.} 
\label{temperaturas0417}
\end{figure}

\begin{figure}
\includegraphics[width=.48\textwidth,angle=0,clip=]{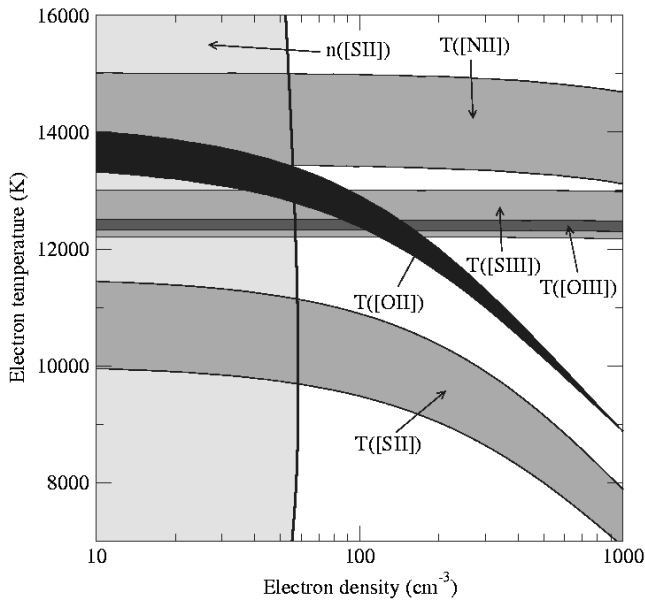}
\caption{Measured line temperatures and electron density for object SDSS
  J162410.11-002202.5. The width of the bands correspond to one $\sigma$ error.} 
\label{temperaturas0364}
\end{figure}

For the three observed objects we have been able to measure four electron
temperatures: T([O{\sc iii}]), T([O{\sc ii}]), T([S{\sc iii}]) and  T([S{\sc
    ii}]). T([N{\sc ii}]) has also been measured in two of the objects. The good
quality of the data has allowed to reach accuracies of the order of 1\% for
T([O{\sc iii}]), 3\% for T([O{\sc ii}]) and 5\% in the case of T([N{\sc ii}]),
T([S{\sc ii}]) and T([S{\sc iii}]). { Figs. \ref{temperaturas0390},
\ref{temperaturas0417} and \ref{temperaturas0364} show the range of the measured
line temperatures and electron density for objects SDSS J002101.03+005248.1, 
SDSS J003218.60+150014.2 and  SDSS J162410.11-002202.5, respectively. The data
for the latter two are very similar, although T([N{\sc ii}]) could not be
measured for SDSS J003218.60+150014.2}. The width of the bands 
correspond to one $\sigma$ error. In the two cases T([O{\sc iii}]) and T([S{\sc
    iii}]) overlap to some extent, while T([O{\sc ii}]) is lower than T([O{\sc
    iii}]) in the first case and slightly higher than T([O{\sc iii}]) in the
second.

The [O{\sc ii}] and [O{\sc iii}] temperatures of the observed objects are shown
in  Fig. \ref{T([OII]-T([OIII])} together with the data of \HII\ galaxies, and
galactic and extragalactic \HII\ regions from
P\'erez-Montero \& D\'\i az (2005) and the photoionization models from
P\'erez-Montero \& D\'\i az (2003) for electron densities n$ _{e} $ = 10, 100
and 500 cm$^{-3} $. The model sequences from Stas\'nska (1980; solid line)
and Stasi\'nska (1990; dashed-dotted line) are also plotted. This figure shows
the effect of the electron density on the derived [O{\sc ii}] temperature which
decreases with increasing density. The data points populate the region of the
diagram spanned  by model sequences but the observational errors are too large
to allow an adequate test of the photoionization models themselves. The
temperatures derived for two of the observed objects:  SDSS J003218.60+150014.2
and SDSS J162410.11-002202.5 lie  close to the theoretical relation  for  an
electron density  of n$ _{e} $= 10 cm $ ^{-3} $,  while for the other one: SDSS
J002101.03+005248.1,  they lie closer to  sequences for electron densities larger
than  100 cm$^{-3}$ . This is consistent with { the trend shown by the values
of} n$ _{e} $ measured for the three objects.  

For this last object, the value of  T([O{\sc ii}]) derived from the measured
T([O{\sc iii}]) according to the widely used fit to the photoionization models by
Stasi\'nska (1990), is larger than the measured one by 2200\,$\degr$K 
which translates into  a lower O$^+$/H$^+$ ionic ratio by a factor of 4 and a
lower total oxygen abundance by 0.17\,dex. It should be remarked that this is
the procedure  usually followed, even in the cases in which  measured values of
T([O{\sc ii}]) exist (see Izotov, Thuan \& Lipovetsky 1994). Also, in doing so,
no uncertainties are attached to the T([O{\sc ii}]) vs.\ T([O{\sc iii}])
relation with the final outcome of a reported T([O{\sc ii}]) which carries only
the usually small observational error found in the derivation of T([O{\sc iii}])
and translates into very small errors in the oxygen ionic and total
abundances. Thus it is possible, and even frequent,  to find reported  values of
T([O{\sc ii}]) and  T([S{\sc iii}]) with quoted fractional errors lower than 
1\% and absolute errors actually less than that quoted for T([O{\sc iii}])
\citep{1998ApJ...500..188I},  and ionic  O$^+$/H$^+$ ratios with errors of only
0.02\,dex. 

Recently, this procedure has been justified by \cite{2006A&A...448..955I} on the
basis of an analysis of SDSS data. In this work it is argued that, despite a
large scatter, the relation between  T([O{\sc ii}]) and T([O{\sc iii}]) derived
from observations follows generally the one obtained by models, and therefore
they adopt T([O{\sc ii}]) as derived from their photoionization models. However,
the large errors attached to the electron temperature determinations, in many
cases around $\pm$\,2000\,$\degr$K for T([O{\sc ii}]), actually precludes the test of such
a statement. In fact, most of the data with the smallest error bars lie below
and above the theoretical relation.  

Regarding the [S{\sc iii}] temperatures, our values fit well the relation
between T([O{\sc iii}]) and T([S{\sc iii}]) found by Garnett (1992) as can be
seen in Fig. \ref{T([SIII]-T([OIII])}. For the objects for which we have measured
T([N{\sc ii}]), this is equal to T([S{\sc iii}]) within the errors. 

The abundances derived for the observed objects show the characteristic low
values found in { strong line \HII\ galaxies
\citep{1991A&AS...91..285T,2006MNRAS.365..454H}}: 
12+log(O/H)\,=\,8.0, within the errors, 
for SDSS J002101.03+005248.1 and SDSS J003218.60+150014.2 and slightly lower
12+log(O/H)\,=\,7.9 for SDSS J162410.11-002202.5.  
The three objects have previous oxygen abundance determinations. They are part
of the first edition of the SDSS \HII\ galaxies with oxygen abundance catalog,
presented by \cite{2004ApJS..153..429K}. These authors derived total oxygen
abundances of 12+log(O/H)\,=\,8.18$\pm$0.04 for SDSS J002101.03+005248.1,
12+log(O/H)\,=\,8.07$\pm$0.02 for SDSS J003218.60+150014.2 and
12+log(O/H\,=\,8.17$\pm$0.01 for SDSS J162410.11-002202.5, higher than ours by
0.17\,dex, but consistent with the values we obtain from the analysis of the SDSS
spectra for two of the objects: SDSS J002101.03+005248.1
(12+log(O/H)\,=\,8.19$\pm$0.03) and SDSS J162410.11-002202.5
(12+log(O/H\,=\,8.14$\pm$0.02). For SDSS J003218.60+150014.2, our analysis of
its SDSS spectrum yields a total oxygen abundance 12+log(O/H)\,=\,7.93$\pm$0.03,
lower than theirs by  0.14\,dex and closer to the value derived by
\cite{2003A&A...397..463U} in their Hamburg/SAO Survey. The spectrum analyzed by
\cite{2004ApJS..153..429K} was extracted from the first SDSS data release. This
might point to a difference in the calibration routines of SDSS spectra from one
release to the other.  


The logarithmic N/O ratios found for the two galaxies for which there are T([O{\sc ii}]) and
T([N{\sc ii}]) determinations are -1.06$\pm$0.10 and -1.26$\pm$0.12. For the
third object we estimate a log(N/O) ratio of -1.11. They point to a constant value
within the errors. It is worth noting that an analysis of the data along the
lines discussed above, \textit{i. e. }  T([O{\sc ii}]) derivation from T([O{\sc
    iii}]) according to Stasi\'nska (1990) models, would provide for one of the
objects, SDSS J002101.03+005248.1, an N/O ratio larger by a factor of 3 (log
N/O\,=\,-0.64). Therefore it is not unplausible that part of the scatter found
in the N/O vs.\ O/H diagram may be due to the methodology employed in the
derivation of the N/O ratios. This is an important effect that should be explored
further.  

Finally, the log(S/O) 
ratios found for the three objects are: -1.83$ \pm $0.16,
-1.44$\pm$0.14 and -1.57$\pm$0.11, barely consistent with solar (S/O\,=\,-1.39)
except for J002101.03+005248.1 which is lower by a factor of about 
2.7. The analysis of the data according to the conventional lines described
above would rise this S/O ratio somewhat, but still keeping it 
below the solar value.

\begin{figure}
\includegraphics[width=.48\textwidth,angle=270]{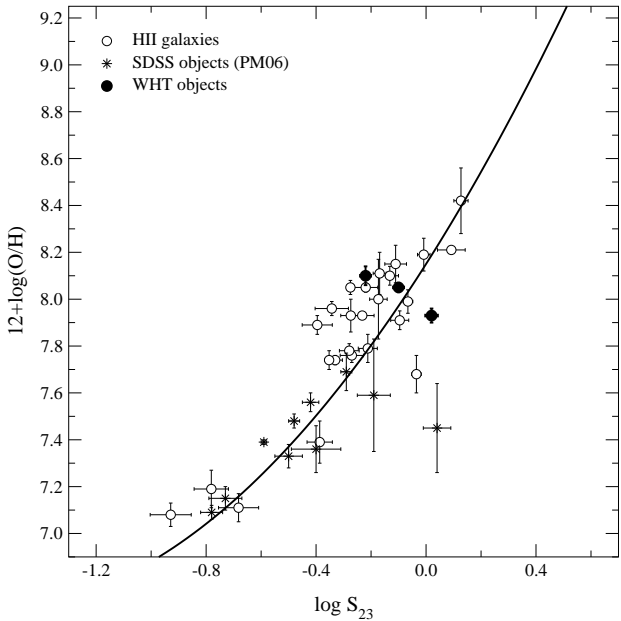}
\caption{{ Relationship between log\,S$_{23}$ and metallicity, represented by
  12+log(O/H), for the observed objects (solid circles), the \HII\ galaxies from
  P\'erez-Montero \& D\'iaz (2005) with data on the [S{\sc
  iii}]\,$\lambda\lambda$\,9069,9532\,\AA\ emission lines (open circles), and
  SDSS BCDs galaxies 
  (asterisks) from   P\'erez-Montero et al.\ 2006. The solid line shows the
  calibration made by P\'erez-Montero \& D\'iaz (2005).}} 
\label{s23-o}
\end{figure}

{ On the other hand, the measurement of the [S{\sc iii}] IR lines allows the
calculation of the sulphur abundance parameter S$ _{23} $
\citep{1996MNRAS.280..720V}. 
\[
S_{23}\,=\,\frac{[S\textrm{\sc ii}]\,\lambda\lambda\,6717,31+[S\textrm{\sc
      iii}]\,\lambda\lambda\,9069,9532}{H\beta}
\]

This parameter constitutes probably the best
empirical abundance indicator for \HII\ 
galaxies, since contrary to what happens for the widely used O$_{23}$ (R$_{23}$)
parameter  \citep{1979MNRAS.189...95P, 1979A&A....78..200A}, the calibration is
linear up to solar abundances, thus solving the degeneration problem usually
presented by this kind of objects. This is particularly dramatic for objects
with log\,O$ _{23}$\,$\geq$\,0.8 and 12+log(O/H)\,$\geq$\,8.0 which can show the
same value of the abundance parameter while having oxygen abundances that differ
by up to an order of magnitude. About 40\,\%\ of the observed \HII\ galaxies
belong to this category \cite[see][]{1999cezh.conf..134D}.
The logarithm of the S$_{23}$ parameter 
derived for our objects are given in the last row of table \ref{abs}. Fig.
\ref{s23-o} shows the points corresponding to  the objects in the log\,S$_{23}$
vs.\ 12+log(O/H) diagram, together with their observational error, and labeled
as WHT objects. The rest of the data correspond to \HII\ galaxies 
from P\'erez-Montero \& D\'iaz (2005) together with the data of ten SDSS BCD
galaxies presented by \cite{2004ApJS..153..429K} and analyzed by
\cite{2006A&A...449..193P} (these objects are labeled as SDSS objects). The
solid line shows the calibration by P\'erez-Montero \& D\'iaz (2005). Three
objects are seen to clearly deviate from this  
line. Two of them correspond to two of the galaxies of \cite{2004ApJS..153..429K} 
for which no data on the [O{\sc ii}]\,3727 line exist and therefore the
O$^+$/H$^+$ ratio is derived from the red [O{\sc ii}]\,7325 lines. Both objects
show low ionization parameters as estimated from the [S{\sc ii}]/[S{\sc iii}]
ratio and relatively high values 
of O/H as derived from the [N{\sc ii}]/H$\alpha$ calibration
\citep*{2002MNRAS.330.69D}. The third object is
Mrk709, an object that shows similar characteristics (see P\'erez-Montero \&
D\'iaz 2003). These objects might be affected by shocks and deserve further
study. Actually the accuracy of the S$_{23}$ calibration for \HII\ galaxies, as
a family, is only 0.10\,dex (see P\'erez-Montero \& D\'iaz 2005). But obviously,
more observations are needed in order to improve the S$_{23}$ calibration and
truly understand the origin of the observed dispersion.}

\subsection{Ionization structure}

The ionization structure of a nebula depends essentially on the shape of the
ionizing continuum and the nebular geometry and can be traced  by the ratio of
successive stages of ionization of the different elements. With our data it is
possible to use the O$^+$/O$ ^{2+} $ and the S$ ^{+} $/S$ ^{2+} $ to probe the
nebular ionization structure. In fact, \cite{1988MNRAS.231..257V} showed that
the quotient of these two quantities that they called ``softness parameter" and
denoted by $ \eta $ is intrinsically related to the shape of the ionizing
continuum and depends on geometry only slightly. If the simplifying assumption
of spherical geometry and constant filling factor is made, the geometrical
effect can be represented by the ionization parameter which, in turn can be
estimated from the [O{\sc ii}]/[O{\sc iii}] ratio. { Actually, the
[O{\sc ii}]/[O{\sc iii}] ratio depends on stellar effective temperature which,
in turn, 
depends on metallicity, in the sense that for a given stellar mass, stars of
higher metallicity have a lower effective temperature.} Since the three observed
objects have similar values of this quotient (between 0.2 and 0.3), { and
show oxygen abundances in a very narrow range}, we can assume they also 
share a common value for their ionization parameter. Under these circumstances,
the value of $ \eta $ points to the temperature of  the ionizing radiation. 

\begin{figure*}
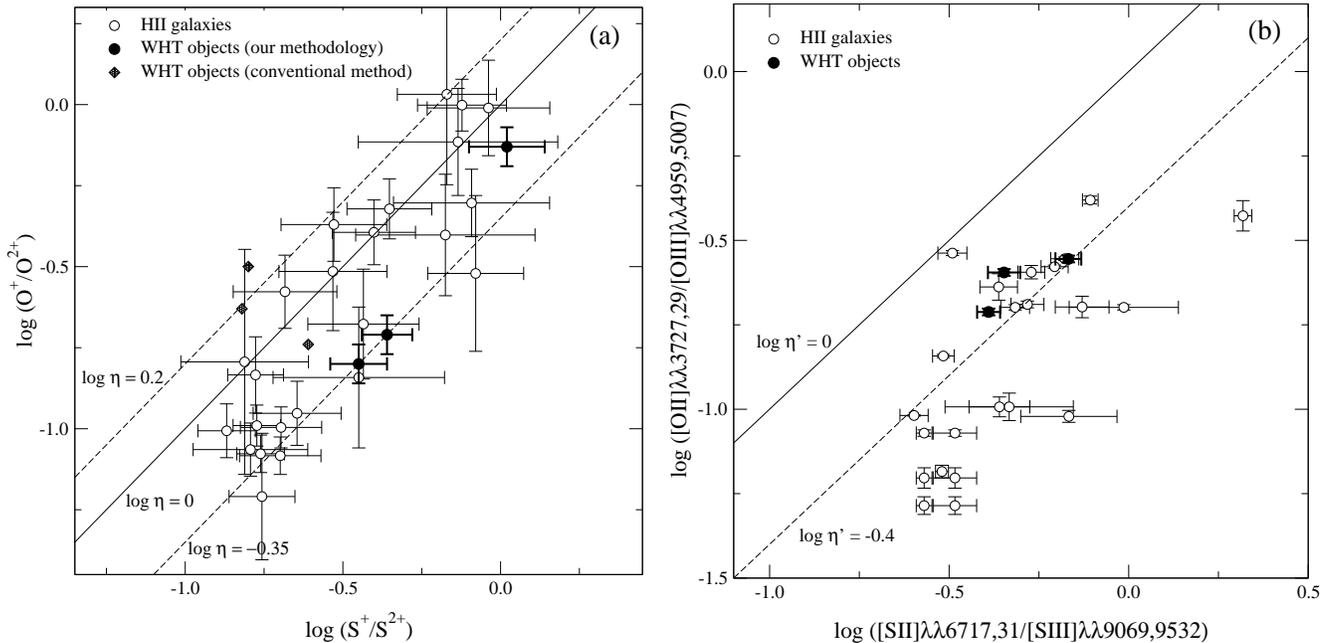

\includegraphics[width=.48\textwidth,angle=0]{figures/eta_SO.eps}\hspace{0.2cm}
\includegraphics[width=.49\textwidth,angle=0]{figures/etaprima_SO.eps}\\
\caption{{ Panel (a): Relation between  log(O$^+$/O$^{2+}$) and
  log(S$^{+}$/S$^{2+}$)  for the observed objects (solid circles for values
  calculated using our methodology and solid diamonds for values derived using
  the conventional method) and the \HII\ galaxies from P\'erez-Montero \& D\'iaz
  (2005)  for which the different oxygen and sulphur ionic ratios can be derived
  (open circles). Diagonals in this  
  diagram correspond to constant values of $\eta$. Panel (b): Relation between
  log([O{\sc ii}]/[O{\sc iii}]) and log([S{\sc ii}]/[S{\sc iii}])  for the
  observed objects (solid circles) and \HII\ galaxies from P\'erez-Montero \&
  D\'iaz (2005) with data on the [S{\sc iii}]\,$\lambda\lambda$\,9069,9532\,\AA\
  emission lines (open circles). Diagonals in this diagram correspond to
  constant values of $\eta$'.}}  
\label{eta}
\end{figure*}

{ In panel (a) of Fig. \ref{eta} we show the relation between
log(O$^+$/O$^{2+}$) and log(S$^{+}$/S$^{2+}$)  for the observed objects (solid
circles for values calculated using our methodology and solid diamonds for
values derived using the conventional method) and the \HII\
galaxies from P\'erez-Montero \& D\'iaz (2005) for which the different oxygen
and sulphur ionic ratios can be derived (open circles).} In this diagram,
diagonal lines correspond to constant values of $ \eta $; the solid line
corresponds to log\,$\eta$\,=\,0.00. The region above this line corresponds to
$\eta$\,$>$\,1 and the region below the line corresponds to $\eta$ $<$ 1. The
dashed lines corresponds to log\,$\eta$\,=\,-0.35 and log\,$\eta$\,=\,0.2 . As
can be seen in the figure, \HII\ galaxies occupy the region of the diagram with
-0.35\,$\leq$\,log\,$\eta$\,$\leq$\,0.20, which corresponds to high values of
the ionizing temperature according to  V\'ilchez \& Pagel (1988). Two of the
observed objects lie on the log\,$\eta$\,=\,-0.35 line. The third one, the one
with the highest excitation, has log\,$\eta$\,=\,-0.15. An analysis along the
conventional lines commented in the  previous subsection, 
would yield the same value of $\eta$ for the highest excitation object (although
with different values of both O$^+$/O$ ^{2+} $ and  S$ ^{+} $/S$ ^{2+} $ ratios)
but widely different values of $\eta$ for the other two objects: 0.19 and 0.30
indicating much lower ionizing temperatures. 
This is inconsistent, however,  with what is
found if the same exercise is performed using the corresponding quotients of
emission lines: { log([O{\sc ii}]/[O{\sc iii}]) vs log([S{\sc ii}]/[S{\sc
iii}])}, { which does not require explicit knowledge of the line
temperatures involved in the derivation of the ionic ratios, and therefore
does not depend on the method to derive or estimate these temperatures. Panel
(b) in Fig. \ref{eta} shows then the purely observational counterpart of
panel (a). } In this diagram diagonal  
lines represent constant values of log\,$\eta$' defined by V\'ilchez \& Pagel
(1988) as:
\begin{eqnarray*}
\lefteqn{log\,\eta{\textrm'}\,=\,log\Big[ \frac{[O{\textrm {\sc
 ii}}]\lambda\lambda\,3727,29\,/\,[O{\textrm {\sc 
 iii}}]\lambda\lambda\,4959,5007}{[S{\textrm {\sc
 ii}}]\lambda\lambda\,6717,31\,/\,[S{\textrm {\sc
 iii}}]\lambda\lambda\,9069,9532}\Big]
{} } \nonumber\\ & & {}
\,=\,log\,\eta - \frac{0.14}{t_e} -0.16   {}
\end{eqnarray*} 
\noindent where t$_e$ is the electron temperature in units of 10$^4$. $\eta$ and
$\eta$' are related through the electron temperature but very weakly, so that a
change in temperature from 7000 to 14000\,$\degr$K implies a change in
log\,$\eta$ by 0.1\,dex, inside observational errors. Also, log\,$\eta$' is
always less than log\,$\eta$. 
In this diagram, the data for the three objects indicate values of log\,$\eta$'
between -0.39 and -0.25, {  consistent with what is found by our analysis but
inconsistent with the results obtained following the conventional
methodology. Metallicity calibrations based on abundances derived according to
this conventional method are probably bound to provide metallicities which are
systematically too high and should therefore be revised.} 


\subsection{The temperature fluctuation scheme.}
\label{variations}

At the end of the 60s and beginning of the 70s, \cite{1967ApJ...150..825P},
\cite{1969BOTT....5....3P}, and \cite{1971BOTT....6...29P} established a
complete analytical formulation to study the discrepancies between the
abundances relative to hydrogen derived from recombination lines (RLs) and
from collisionally excited lines (CELs) when a constant electron temperature 
is assumed. In the first of these 
works, Peimbert proposed that this discrepancy is due to spatial temperature
variations which can be characterized by two parameters: the average
temperature weighted by the square of the density over the volume considered,
T$_0$, and the root mean square temperature fluctuation, t$^2$. They are given by  
\begin{eqnarray}
\label{t0}
T_0(X^{i+})\,=\,\frac{\int T_e\,N_e\,N(X^{i+})\,dV}{\int N_e\,N(X^{i+})\,dV}
\end{eqnarray}
\noindent
and
\begin{eqnarray}
\label{t2}
t^2(X^{i+})\,=\,\frac{\int(T_e-T_0(X^{i+}))^2\,N_e\,N(X^{i+})\,dV}{T_0(X^{i+})^2\int N_e\,N(X^{i+})\,dV}
\end{eqnarray}
\noindent
where N$_e$ and N$(X^{+i}$) are the local electron and ion densities
of the observed emission lines, respectively;
T$_e$ is the local electron temperature; and $V$ is the observed volume
\citep{1967ApJ...150..825P}.  

It is possible to obtain the values of T$_0$ and t$^2$ using different
methods. One possibility is to compare the electron temperatures obtained using
two independent ways. Generally, temperatures originated in different zones of
the nebula are used, one  that represents the warmest regions and another
representative of the coldest ones. In the absence of  temperatures derived from
recombination lines, the  temperatures estimated from the hydrogen
discontinuities, either from Balmer or Paschen series, representative of the
temperature of the neutral gas, can be used.

We have followed 
\cite{2000ApJ...541..688P,2002ApJ...565..668P,2004ApJS..150..431P} and
\cite{2003ApJ...595..247R} to derive the values of T$_0$ and t$^2$ for our WHT
spectra by combining the Balmer temperature, T(Bac), and the
temperature derived from the collisional [O{\sc iii}] lines,
T([O{\sc iii}]), therefore assuming a simple one-zone ionization scheme. Then, the
relation for these two temperatures is given by:

\begin{eqnarray}
\label{toiii}
T([O\textrm{\sc iii}])\,=\,T_0\,\Big[1+\frac{1}{2}\Big(\frac{91300}{T_0}-3\Big)\,t^2\Big]
\end{eqnarray}
\noindent
and
\begin{eqnarray}
\label{tbac}
T(Bac)\,=\,T_0\big(1-1.67t^2\big)
\end{eqnarray}

\noindent
The solution of this system of equations for each galaxy along with their
corresponding errors are listed in table \ref{fluc}.
The temperature fluctuations are almost negligible for two of our
objects. They have t$^2$-values very similar to the ones derived by
\cite{2003ApJ...592..846L} by combining observations from the literature and
photoionization  models for some BCDs and extragalactic \HII\ regions. 
{ \cite{2006astro.ph..3134G} used the Balmer and Paschen jumps to determine the
temperatures of the H$^+$ zones of 22 low-metallicity \HII\ regions in 18 BCD
galaxies, one extragalactic \HII\ region in M101 and 24 \HII\ emission-line
galaxies selected from the DR3 of the SDSS. They found that these temperatures
do not differ, in a statistical sense, from the temperatures of the [O{\sc iii}]
zones, given t$^2$-values are close to zero.
The greater t$^2$-value obtained for SDSS J003218.60+150014.2 is in the range of
the values derived for giant extragalactic \HII\ regions 
\cite[see][]{1994ApJ...437..239G, 2005ApJ...634.1056P,2005A&A...444..723J} or
\HII\ regions in the Magellanic Clouds, such as 30 Doradus, LMC\,N11B and
SMC\,N66 \citep{2003MNRAS.338..687T}.} 

\setlength{\minrowclearance}{2pt}

\begin{table}
\caption{T$_0$ and t$^2$ parameters}
\label{fluc}
\begin{center}
\begin{tabular}{l c c }
\hline
\hline
       name               &     T$_0$         &   t$^2$         \\
\hline
SDSS J002101.03+005248.1  &    1.24$\pm$0.35  &   0.004$^{+0.044}_{-0.004}$  \\[3pt]
SDSS J003218.60+150014.2  &    1.08$\pm$0.21  &   0.066$\pm$0.026  \\[2pt]
SDSS J162410.11-002202.5  &    1.24$\pm$0.30  &   0.001$^{+0.037}_{-0.001}$  \\[2pt]
\hline

\multicolumn{3}{l}{T$_0$ in 10$^4$\,$\degr$K. Note that t$^2$ is always greater than zero.}

\end{tabular}
\end{center}
\end{table}

In the limit of low densities and small optical depths, and for t$^2$ much lower
than one, the electronic temperature for
helium, T(He{\sc ii}), is proportional to $\langle\alpha\rangle$ and $\beta$,
the average value of the power of the temperature for the helium lines used to
calculate the ionic abundances of He$^+$,  and the corresponding one  for H$\beta$,
respectively. Then, for $\langle\alpha\rangle$ different from $\beta$, 
this temperature is given by equation (14) of Peimbert (1967):
\begin{eqnarray}
\label{the}
T(He\textrm{\sc ii})\,=\,T(He\textrm{\sc ii},H\textrm{\sc
  ii})\,=\,T_0\Big[1+\big(\langle\alpha\rangle+\beta-1\big)\frac{t^2}{2}\Big] 
\end{eqnarray}

The value of the power of the temperature for each helium line in the
low density limit has been obtained from \cite{1999ApJ...514..307B}. We have
calculated $\langle\alpha\rangle$  as the average value of $\alpha$ weighted
according to the observational errors \citep{2000ApJ...541..688P}. We have
obtained values of $\langle\alpha\rangle$ equal to -1.37, -1.42, and -1.43 for
SDSS J002101.03+005248.1, SDSS J003218.60+150014.2, and SDSS
J162410.11-002202.5, respectively. The value of $\beta$ has been obtained from
\cite{1995MNRAS.272...41S} and is equal to -0.89. The results for T(He{\sc ii})
and their corresponding errors are listed in the last row of table \ref{temden}.

The temperature for H$\beta$, T(H$\beta$), can be calculated from equation (20)
in \cite{1969BOTT....5....3P} as:  

\begin{eqnarray}
\label{thbeta}
T(H\beta)=T_0\Big[1-0.95\,t^2\Big]
\end{eqnarray}

\noindent where we have taken $\beta$ = -0.89 as above. 
The derived values for T(H$\beta$) and their errors are also given in table
\ref{temden}.

The  line temperature for a collisionally excited line, CEL, for t$^2\ll1$, ($\Delta
E_{CEL}$/kT$_0$-1/2)\,$\neq$\,0, and $\alpha$\,$\neq$\,0 is given by equation
(20) of \cite{1969BOTT....5....3P}: 

\begin{eqnarray}
\label{tcel}
\lefteqn{T_{CEL}\,=\,T_0\,\Big\{1
{} } \nonumber\\ & & {} 
+\Big[\frac{(\Delta E_{CEL}/kT_0)^2-3\Delta E_{CEL}/kT_0+3/4}{\Delta E_{CEL}/kT_0-1/2}\Big]\frac{t^2}{2}\Big\}  {}
\end{eqnarray}

\noindent where  $\Delta E_{CEL}$\,=\,$\Delta E_{mn}$ is the energy
difference, in eV, between the upper (m) and lower (n)
levels respectively of the atomic transition that produces the line.

For the case of t$ ^{2} > $ 0, and assuming a one-zone ionization scheme,
we can derive the ionic abundances using the values calculated for
t$^2$ equal to zero, and equation (15) of \cite{2004ApJS..150..431P}:

\begin{eqnarray}
\label{abunt2}
\lefteqn{\Big[\frac{N(X^{+i})}{N(H^{+})}\Big]_{t^2>0}\,=\,
  \frac{T(H\beta)^{-0.89}T(\lambda_{mn})^{0.5}}{T([O\textrm{\sc iii}])^{-0.37}}
   } \nonumber\\ & & 
  {} \times\, exp\Big[-\frac{\Delta E}{k\,T([O\textrm{\sc iii}])}+\frac{\Delta
  E}{k\,T(\lambda_{mn})}\Big]\,
  \nonumber\\ & & 
  {} \times\, \Big[\frac{N(X^{+i})}{N(H^{+})}\Big]_{t^2=0} {} 
\end{eqnarray}

\noindent where T($\lambda_{mn}$) is given by equation (\ref{tcel}).
Using this expression, we have calculated the effect of the temperature
fluctuations on the O$ ^{2+} $/H$ ^{+}$ abundance. In fact, given the high
excitation of the object, this ionic abundance carries the highest weight in the
total abundance of oxygen.  The recalculated value  is 12+log(O$ ^{2+}
$/H$^{+}$)\,=\,8.09$\pm$0.12 which yields a  total oxygen abundance
12+log(O/H)\,=\,8.12$\pm$0.11. These values are  higher than those given in
Table \ref{abs} by 0.23 and 0.22\,dex respectively.  
\medskip

\section{Conclusions}

We have performed a detailed analysis of newly obtained spectra of three 
\HII\ galaxies
selected from the Sloan Digital Sky Survey Data Release 2, covering from 3200 to
10550 \AA\ in wavelength. For the three objects we have measured four line
temperatures: T([O{\sc iii}]), T([S{\sc iii}]), T([O{\sc ii}]) and T([S{\sc
    ii}]) and the Balmer continuum temperature T(Bac). For two of the objects we
have also measured  T([N{\sc ii}]). These measurements and a careful and
realistic treatment of the observational errors yield total oxygen abundances
with accuracies between 5 and 9\%. The fractional error is as low as 2\% for the
ionic O$ ^{2+} $/H$ ^{+} $ ratio due to the small errors associated with the
measurement of  the strong nebular lines of [O{\sc iii}] and the derived
T([O{\sc iii}]),
but increases to 15\% for the O$^{+}$/H$^{+}$ ratio. The accuracies are lower
in the case of the abundances of sulphur (of the order of 20\% for S$^+$ and
12\% for S$^{2+}$) due to the presence of larger observational errors both in
the measured line fluxes and the derived electron temperatures. The error
increases further (up to 30\%) for the total abundance of sulphur due to the
uncertainties in the ICF.  

This is in contrast with the small errors quoted for line temperatures other
than T([O{\sc iii}]) in the literature, in part due to the commonly assumed
methodology of deriving them from the measured T([O{\sc iii}]) through
theoretical relations. These relations are found from photoionization models and
no uncertainty is attached to them; therefore, the obtained line temperatures
carries only the observational error found for the T([O{\sc iii}])
measurement. If this methodology were to be adopted, the theoretical relations
should be adequately tested and empirical relations between the relevant line
temperatures should be obtained in order to quantify the corresponding model
uncertainties.  

We have compared our obtained spectra with those downloaded from the SDSS DR3
finding a satisfactory agreement. The analysis of these spectra yields values of
line temperatures and elemental ionic and total abundances which are in general
agreement with those derived from the WHT spectra, although for most quantities,
they can only be taken as estimates since, due to the lack of direct
measurements of the required lines, theoretical models had to be used whose
uncertainties are impossible to quantify. Unfortunately, the spectral coverage
of SDSS precludes the simultaneous observation of the [O{\sc
    ii}]\,$\lambda\lambda$\,3727,29\,\AA\ and [S{\sc
    iii}]\,$\lambda\lambda$\,9069, 9532\,\AA\ lines, and therefore the analysis
can never be complete.

The ionization structure found for the observed objects from the O$^{+} $/O$
^{2+} $ and S$^{+} $/S$ ^{2+} $ ratios points to high values of the ionizing
radiation as traced by the values of the ``softness parameter" $\eta$ which is
less than one for the three objects. Line temperatures derived from
T([O{\sc iii}]) with the use of current photoionization models yield for the two
highest excitation objects,  much higher values of $\eta$ which would imply
lower ionizing temperatures. This is however inconsistent with the ionization
structure as probed by the measured emission line intensities. 

Finally, we have measured the Balmer continuum temperature for the three
observed objects and derived the temperature fluctuations as defined by
\cite{1967ApJ...150..825P}. Only for one of the objects, the temperature
fluctuation is significant and could lead to higher oxygen abundances by about
0.20\,dex.

\section*{Acknowledgements}

We are grateful to Jorge Garc\'\i a-Rojas and C\'esar Esteban  for calculating
the effects of the temperature fluctuations over the derived
ionic and total abundances. We are also grateful to an anonymous referee
for his/her careful and constructive revision of the manuscript.

The WHT is operated in the island of La Palma by the Isaac Newton Group
in the Spanish Observatorio del Roque de los Muchachos of the Instituto
de Astrof\'\i sica de Canarias. We thank the Spanish allocation committee
(CAT) for awarding observing time.

Funding for the creation and distribution of the SDSS Archive has been provided
by the Alfred P. Sloan Foundation, the Participating Institutions, the National
Aeronautics and Space Administration, the National Science Foundation, the US
Department of Energy, the Japanese Monbukagakusho, and the Max Planck
Society. The SDSS Web site is http://www.sdss.org.

The SDSS is managed by the ARC for the Participating Institutions. The
Participating Institutions are the University of Chicago, Fermilab, the
Institute for Advanced Study, the Japan Participation Group, The Johns Hopkins
University, the Korean Scientist Group, Los Alamos National Laboratory, the Max
Planck Institute for Astronomy (MPIA), the Max Planck Institute for Astrophysics
(MPA), New Mexico State University, the University of Pittsburgh, the University
of Portsmouth, Princeton University, the United States Naval Observatory, and
the University of Washington.

This research has made use of the NASA/IPAC Extragalactic Database (NED) which
is operated by the Jet Propulsion Laboratory, California Institute of
Technology, under contract with the National Aeronautics and Space
Administration. 

This research has made use of the SIMBAD database, operated at CDS, Strasbourg,
France.

This work has been partially supported by DGICYT project AYA-2004-02860-C03 and
the Spanish MEC FPU grant AP2003-1821. Elena Terlevich acknowledges support from 
the Spanish MEC through a grant for a sabbatical visit SAB-2004-0148. 
She and Roberto would like to thank the hospitality of the Astrophysics Group of
the UAM during the completion of this work.

\bibliographystyle{mn2e}
\bibliography{HIIgal-WHT-Hagele}
\end{document}